\documentclass[preprint2]{aastex61}

\received{}
\revised{}
\accepted{}
\submitjournal{ApJ}

\shorttitle{Misclassified B stars in the {\it Kepler} field}
\shortauthors{Zhang et al.}

\begin{document}

\title{Misclassified B stars in the {\it Kepler} field}

\correspondingauthor{Chunguang Zhang}
\email{cgzhang@nao.cas.cn}

\author{Chunguang Zhang}
\affiliation{Key Laboratory of Optical Astronomy, National Astronomical Observatories, Chinese Academy of Sciences, Beijing 100101, China}

\author{Chao Liu}
\affiliation{Key Laboratory of Optical Astronomy, National Astronomical Observatories, Chinese Academy of Sciences, Beijing 100101, China}

\author{Yue Wu}
\affiliation{Key Laboratory of Optical Astronomy, National Astronomical Observatories, Chinese Academy of Sciences, Beijing 100101, China}

\author{Yangping Luo}
\affiliation{Physics and Space Science College, China West Normal University, Nanchong 637002, China}

\author{Xiaobin Zhang}
\affiliation{Key Laboratory of Optical Astronomy, National Astronomical Observatories, Chinese Academy of Sciences, Beijing 100101, China}

\author{Licai Deng}
\affiliation{Key Laboratory of Optical Astronomy, National Astronomical Observatories, Chinese Academy of Sciences, Beijing 100101, China}

\author{Jianning Fu}
\affiliation{Department of Astronomy, Beijing Normal University, Beijing 100875, China}

\author{Yong Zhang}
\affiliation{Nanjing Institute of Astronomical Optics \& Technology, National Astronomical Observatories, Chinese Academy of Sciences, Nanjing 210042, China}

\author{Yonghui Hou}
\affiliation{Nanjing Institute of Astronomical Optics \& Technology, National Astronomical Observatories, Chinese Academy of Sciences, Nanjing 210042, China}

\author{Yuefei Wang}
\affiliation{Nanjing Institute of Astronomical Optics \& Technology, National Astronomical Observatories, Chinese Academy of Sciences, Nanjing 210042, China}

\begin{abstract}
Stellar fundamental parameters are important in the asteroseismic study of {\it Kepler} light curves. However, the most used estimates in the {\it Kepler} Input Catalog (KIC) are not accurate enough for hot stars. Using a sample of B stars from the Large Sky Area Multi-Object Fiber Spectroscopic Telescope (LAMOST) spectral survey, we confirmed the systematic underestimation in KIC effective temperature and overestimation in KIC surface gravity. The good agreement between LAMOST and other follow-up observations proved the accuracy of effective temperature and surface gravity of B stars derived from LAMOST low-resolution spectra. By searching through LAMOST data, we found four misclassified main-sequence B stars in the {\it Kepler} field, which had been previously classified as A-type variables. We present spectroscopic and detailed frequency analysis of these four stars based on LAMOST spectra and {\it Kepler} photometry.
\end{abstract}

\keywords{stars: early-type --- stars: fundamental parameters --- stars: oscillations}

\section{Introduction}\label{sec1}

With the extraordinary photometric precision and near-continuous observations, the {\it Kepler} mission \citep{Kepler} has revolutionized the study of exoplanets and stellar physics. Before the loss of a second reaction wheel, {\it Kepler} had monitored more than $150,000$ stars for 4 years in the originally fixed field of view (FoV). The collected data enable one to study in unprecedented detail the internal structures of stars by means of asteroseismology \citep{DeRidder2009, Papics2014, Keen2015, Kurtzetal2015, Saio2015, Schmid2015, VanReeth2015}.

{\it Kepler} has revealed some surprising results of the classical pulsators along the main sequence (MS), such as $\gamma$ Dor stars, $\delta$ Sct stars, slowly pulsating B (SPB) stars, and $\beta$ Cep stars. The distribution of oscillating A and F stars extend well beyond the instability strips defined by ground-based observations and theoretical calculations, and hybrid pulsators are far more common than previously expected \citep{GABetal2010, UMGetal2011}. There are also non-pulsating stars within the instability strips \citep{BD2011, Balonaetal2011}.

Further investigations of these problems require reliable classification and statistics of different types of variable stars. However, it is difficult to discriminate between $\gamma$ Dor and SPB stars, or between $\delta$ Sct and $\beta$ Cep stars based solely on their oscillations, because the period ranges of these different types of stars overlap. Therefore accurately determined stellar parameters, such as effective temperature $T_\mathrm{eff}$, surface gravity $\log g$, and metallicity, are crucial for the identification of MS oscillations and precision asteroseismology. However, only a small fraction of these upper MS stars in the {\it Kepler} field have parameters derived from ground-based spectroscopic or photometric follow-up observations. Most of the classification and statistical analysis of variable stars still rely on the {\it Kepler} Input Catalog \citep[KIC;][]{KIC}. \citet{PAMZetal2012} found that the KIC systematically underestimates $T_\mathrm{eff}$. The underestimation increases toward high temperature and becomes rather significant for B stars \citep{Balonaetal2011, LTSetal2011, MNJMK2012, TLSU2013}. As warned by \citet{KIC}, ``the KIC $T_\mathrm{eff}$ estimates are untrustworthy for $T_\mathrm{eff} \ge 10^4$ K" due to the lack of $u$-band data.

Underestimated $T_\mathrm{eff}$ may lead to misidentification of B stars. In the case of oscillation, a $\beta$ Cep (or SPB) star could easily be misidentified as a $\delta$ Sct (or $\gamma$ Dor) candidate. This may further affect our conclusions on the distribution and proportion of different types of stellar oscillations. Therefore in order to fully exploit the large amount of {\it Kepler} data, it is desirable to obtain accurate stellar parameters efficiently from ground-based follow-up observations.

The Large Sky Area Multi-Object Fiber Spectroscopic Telescope \citep[LAMOST, also known as Guoshoujing Telescope;][]{LAMOST} has the potential to fulfill this purpose. LAMOST is a reflecting Schmidt telescope with a large aperture (4 m) and a wide FoV ($5^\circ$). The $4000$ fibers on the focal plane enable it to collect low-resolution ($R=1800$) spectra of thousands of objects down to magnitude 19 in a single exposure \citep{Lsurvey}. The LAMOST--{\it Kepler} project aims to determine the parameters of more than 1 million objects in the {\it Kepler} field \citep{DeCat2015}. Using $12,000$ stars, \citet{DZZetal2014} confirmed the systematic underestimation of metallicity in KIC, and gave a correction relation based on LAMOST metallicity. In this paper, we verify the discrepancy in B-star parameters between the KIC and ground-based observations using LAMOST data in Section \ref{sec2}, and report the identification of four misclassified MS B stars in Section \ref{sec3}. The oscillation properties of the four stars are studied in detail based on four years of {\it Kepler} observation in Section \ref{sec4}. In the last section, we present our conclusions.

\section{Comparison of stellar parameters}\label{sec2}

We have collected the published values of $T_\mathrm{eff}$ and $\log g$ for {\it Kepler} MS B stars, including those from high-resolution spectroscopy by \citeauthor{LTSetal2011} (2011; $R=32,000$), \citeauthor{BM2011} (2011; $R=15,000$--$80,000$), \citeauthor{TLSU2013} (2013; $R=32,000$), and \citeauthor{Papics2013} (2013; $R=85,000$), and low-resolution spectroscopy by \citeauthor{Balonaetal2011} (2011; $R=550$). \citet{Balonaetal2011} also derived $T_\mathrm{eff}$ and $\log g$ for 38 B stars from Str\"omgren photometry. We searched for these stars in LAMOST Data Release 2 (DR2)\footnote[1]{http://dr2.lamost.org}, and made comparisons with KIC and the follow-up observations.

LAMOST DR2 comprises all of the observations before 2014 June, including the pilot survey and the first two years of the regular survey. More than $56,000$ spectra of {\it Kepler} targets have been collected. The stellar parameters of OBA stars are determined using the ULySS package\footnote[2]{http://ulyss.univ-lyon1.fr} \citep{KPBW2009, WSPGK2011}, consisting of minimizing the $\chi^2$ value between an observation and a model spectrum via full spectral fitting within 3900--5700{\r A} wavelength range (avoiding the combining section of the blue-arm and red-arm spectra). The ELODIE library \citep{ELODIE} was used as a reference. For normal FGK-type stars, their atmospheric parameters are determined using the LAMOST stellar parameter pipeline \citep{Lpipeline, Wu2014, Ren2016}. The estimation of stellar parameters is sensitive to the signal-to-noise ratio (S/N) of the spectra \citep{Lpipeline}. In our selection of B stars, we employed a criterion of the S/N in $g$ band $\ge 50$.

By cross identification, we found $18$ {\it Kepler} MS B stars in LAMOST DR2 with other spectroscopic or photometric follow-up observations. The KIC numbers and stellar parameters from KIC, LAMOST, and the literature of these stars are listed in Table~\ref{tab1}. \citet{Papics2013} identified KIC 4931738 as a double-lined spectroscopic binary system consisting of two MS B stars. The $T_\mathrm{eff}$ of KIC 4931738 in Table~\ref{tab1} is that of the primary derived from disentangled spectra. The errors in stellar parameters from LAMOST and high-resolution spectroscopy are also given in Table~\ref{tab1}. External calibrations show that the average uncertainties in LAMOST $T_\mathrm{eff}$, $\log g$, and [Fe/H] of OBA stars are $4.3\%$, 0.18, and 0.15, respectively \citep{MILES, WSPGK2011}. For the KIC, the typical errors in $T_\mathrm{eff}$ and $\log g$ are 200\,K and 0.5, respectively \citep{LTSetal2011}.  For low-resolution spectroscopy, the formal fitting errors in $T_\mathrm{eff}$ and $\log g$ are 200\,K and 0.05, respectively \citep{Balonaetal2011}. The errors in parameters derived from Str\"omgren photometry were not provided \citep{Balonaetal2011}, but the calibration method \citep{Balona1994} gives typical errors of $5\%$ in $T_\mathrm{eff}$ and 0.1 in $\log g$.

\begin{deluxetable*}{ccclcrllcc}
\rotate
\tablewidth{0pt}
\tablecaption{Comparison of stellar parameters of {\it Kepler} B stars.\label{tab1}}
\tablehead{
\colhead{KIC Number} & \multicolumn{2}{c}{KIC} & \multicolumn{3}{c}{LAMOST} & \multicolumn{2}{c}{Spectroscopy} & \multicolumn{2}{c}{Photometry}\\
& \colhead{$T_\mathrm{eff}$ (K)} & \colhead{$\log g$} & \colhead{$T_\mathrm{eff}$ (K)} & \colhead{$\log g$} & \colhead{[Fe/H]} & \colhead{$T_\mathrm{eff}$ (K)} & \colhead{$\log g$} & \colhead{$T_\mathrm{eff}$ (K)} & \colhead{$\log g$}}
\startdata
\phn3629496 & \phn9796 &   4.512 & 11603\,$\pm$\,45  & 3.99\,$\pm$\,0.04 & -0.02\,$\pm$\,0.02 & 11320\,$\pm$\,210\,$^a$  & 3.75\,$\pm$\,0.10\,$^a$ &    \nodata     &    \nodata \\
\phn3756031 &    11177 &   4.241 & 15869\,$\pm$\,73  & 3.69\,$\pm$\,0.06 &  0.01\,$\pm$\,0.02 & 15980\,$\pm$\,310\,$^b$  & 3.75\,$\pm$\,0.06\,$^b$ &    16310\,$^c$ & 4.19\,$^c$ \\
\phn3839930 &    11272 &   4.277 & 15407\,$\pm$\,63  & 3.70\,$\pm$\,0.05 & -0.09\,$\pm$\,0.03 & 16500\,$^c$              & 4.2\,$^c$               &    17160\,$^c$ & 4.51\,$^c$ \\
\phn3865742 &    11406 &   4.355 & 19007\,$\pm$\,144 & 3.87\,$\pm$\,0.07 & -0.06\,$\pm$\,0.04 & 19500\,$^c$              & 3.7\,$^c$               &    20190\,$^c$ & 4.20\,$^c$ \\
\phn4136285 &  \nodata & \nodata & 15968\,$\pm$\,55  & 3.68\,$\pm$\,0.05 &  0.12\,$\pm$\,0.02 & 16000\,$\pm$\,1000\,$^d$ & 4.0\,$\pm$\,0.1\,$^d$   &    \nodata     &    \nodata \\
\phn4931738 &    10136 &   4.420 & 12726\,$\pm$\,115 & 4.05\,$\pm$\,0.10 &  0.03\,$\pm$\,0.06  & 13730\,$\pm$\,200\,$^e$  & 3.97\,$\pm$\,0.05\,$^e$ &    \nodata     &    \nodata \\
\phn5130305 & \phn9533 &   4.144 & 10497\,$\pm$\,69  & 3.91\,$\pm$\,0.08 & -0.34\,$\pm$\,0.05  & 10670\,$\pm$\,200\,$^b$  & 3.86\,$\pm$\,0.07\,$^b$ &    10190\,$^c$ & 4.38\,$^c$ \\
\phn7599132 &    10251 &   3.624 & 10687\,$\pm$\,43  & 3.94\,$\pm$\,0.04 & -0.30\,$\pm$\,0.03  & 11090\,$\pm$\,140\,$^b$  & 4.08\,$\pm$\,0.06\,$^b$ &    10560\,$^c$ & 4.39\,$^c$ \\
\phn8057661 & \phn8230 &   3.974 & 20063\,$\pm$\,126 & 3.84\,$\pm$\,0.04 & -0.11\,$\pm$\,0.03  & \nodata                  & \nodata                 &    21360\,$^c$ & 4.23\,$^c$ \\
\phn8177087 & \phn9645 &   4.104 & 14909\,$\pm$\,86  & 3.41\,$\pm$\,0.07 &  0.04\,$\pm$\,0.03  & 13330\,$\pm$\,220\,$^b$  & 3.42\,$\pm$\,0.06\,$^b$ &    13380\,$^c$ & 3.79\,$^c$ \\
\phn8381949 & \phn9782 &   4.394 & 20402\,$\pm$\,90  & 3.74\,$\pm$\,0.05 & -0.08\,$\pm$\,0.02  & 24500\,$^c$              & 4.3\,$^c$               &    21000\,$^c$ & 3.82\,$^c$ \\
\phn8389948 & \phn8712 &   3.612 & 10049\,$\pm$\,52  & 3.77\,$\pm$\,0.07 & -0.43\,$\pm$\,0.04  & 10240\,$\pm$\,340\,$^b$  & 3.86\,$\pm$\,0.12\,$^b$ & \phn9690\,$^c$ & 4.33\,$^c$ \\
\phn8714886 & \phn9142 &   4.093 & 16285\,$\pm$\,64  & 3.76\,$\pm$\,0.05 & -0.10\,$\pm$\,0.02  & 19000\,$^c$              & 4.3\,$^c$               &    18505\,$^c$ & 4.49\,$^c$ \\
\phn9964614 & \phn8915 &   4.067 & 19109\,$\pm$\,80  & 3.69\,$\pm$\,0.03 & -0.01\,$\pm$\,0.02  & 20300\,$^c$              & 3.9\,$^c$               &    19471\,$^c$ & 3.75\,$^c$ \\
   10536147 &    12490 &   5.890 & 20605\,$\pm$\,97  & 3.78\,$\pm$\,0.05 & -0.03\,$\pm$\,0.02  & 20800\,$^c$              & 3.8\,$^c$               &    \nodata     &    \nodata \\
   10658302 &    14809 &   6.076 & 15179\,$\pm$\,91  & 3.75\,$\pm$\,0.07 & -0.02\,$\pm$\,0.05  & 15900\,$^c$              & 3.9\,$^c$               &    \nodata     &    \nodata \\
   10960750 &  \nodata & \nodata & 20219\,$\pm$\,45  & 3.80\,$\pm$\,0.02 & -0.11\,$\pm$\,0.02  & 19960\,$\pm$\,880\,$^b$  & 3.91\,$\pm$\,0.11\,$^b$ &    20141\,$^c$ & 3.85\,$^c$ \\
   11360704 &    12400 &   4.934 & 17644\,$\pm$\,78  & 3.70\,$\pm$\,0.07 & -0.10\,$\pm$\,0.02  & 20700\,$^c$              & 4.1\,$^c$               &    17644\,$^c$ & 3.89\,$^c$ \\
\enddata
\tablerefs{$^a$\,Tkachenko et al. (2013), $^b$\,Lehmann et al. (2011), $^c$\,Balona et al. (2011), $^d$\,Bohlender \& Monin (2011), $^e$\,P\'apics et al. (2013).}
\end{deluxetable*}

\begin{figure}[htbp]
\epsscale{}
\plotone{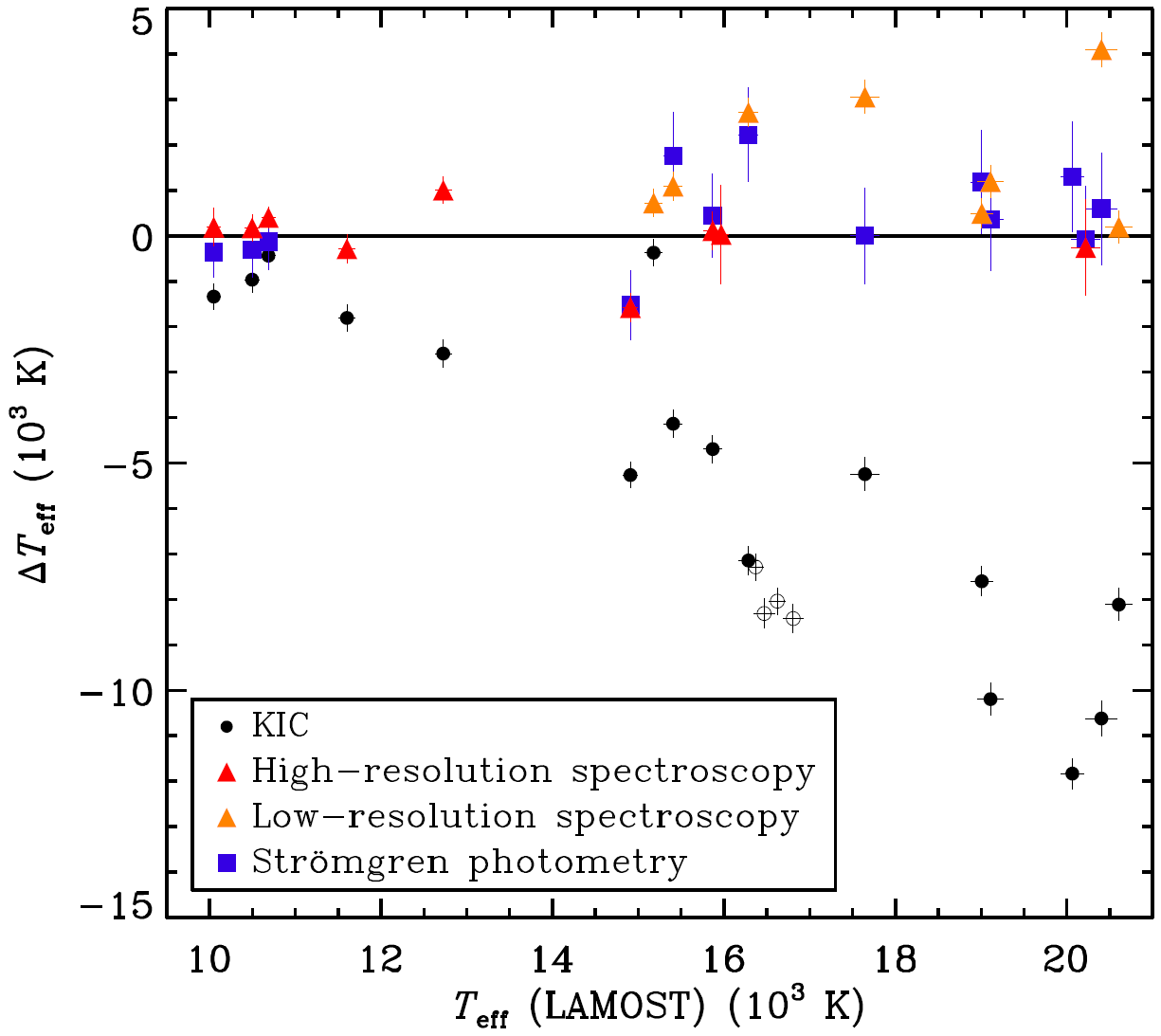}
\caption{Difference in effective temperature between other observations and LAMOST as a function of LAMOST $T_\mathrm{eff}$. The open circles show $\Delta T_\mathrm{eff}$ between KIC and LAMOST for the four B stars reported in this paper.}
\label{fig1}
\end{figure}

Figure~\ref{fig1} shows the temperature difference $\Delta T_\mathrm{eff}$ between other sources and LAMOST. As can be seen, LAMOST values are in good agreement with spectroscopic and photometric data, especially with those from high-resolution spectroscopy. The KIC estimates are clearly too low. The underestimation increases almost linearly with increasing $T_\mathrm{eff}$, and becomes more than $10,000$\,K at high-temperature end. In the work of Lehmann et al. (2011), $\Delta T_\mathrm{eff}$ was fitted using a second-order polynomial with the zero point set at $7,000$\,K. Such underestimation in $T_\mathrm{eff}$ could result in the misclassification of B stars. Three stars in Table~\ref{tab1} with $T\mathrm{_{eff}(KIC)}<10,000$\,K (KIC 3629496, KIC 5130305, and KIC 8389948) were indeed cataloged as A stars \citep{Balona2013}.

\begin{figure}[htbp]
\epsscale{}
\plotone{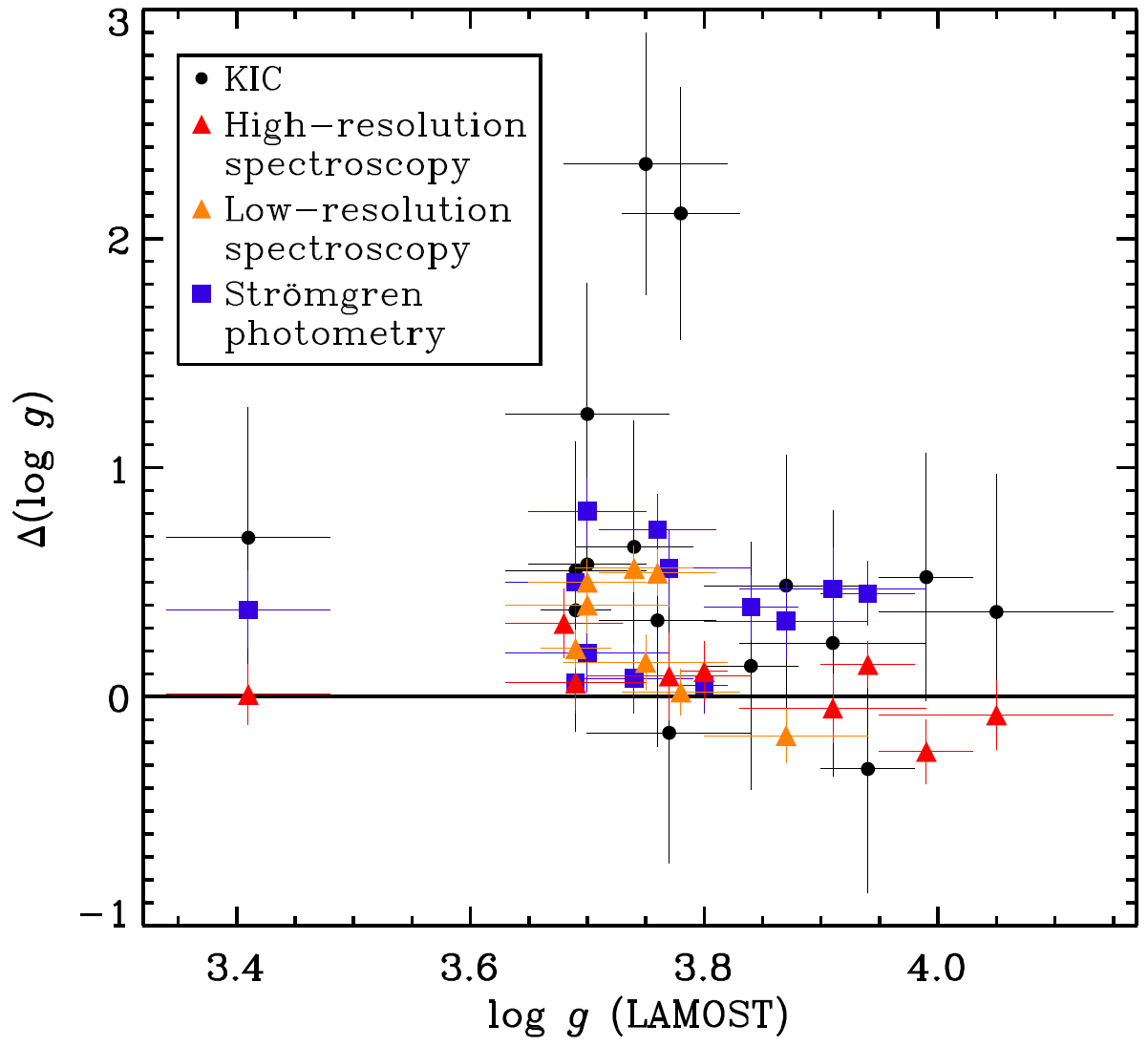}
\caption{Difference in surface gravity between other observations and LAMOST as a function of LAMOST $\log g$.}
\label{fig2}
\end{figure}

The surface gravity difference $\Delta(\log g)$ between other observations and LAMOST is shown in Figure~\ref{fig2}. Although the results from low-resolution spectroscopy and Str\"omgren photometry seem more close to the KIC values, comparison between LAMOST and the KIC shows that KIC $\log g$ is systematically higher, confirming the finding of \citet{LTSetal2011} based on high-precision spectroscopy. Considering the large error in KIC $\log g$, the difference in $\log g$ is less significant than that in $T_\mathrm{eff}$. However, for two stars, namely KIC 10536147 and KIC 10658302, KIC $\log g$ does show clear overestimation, which is large enough for them to be misclassified as compact objects. In fact, KIC 10658302 was selected as a candidate of compact stellar objects based on KIC data, but was then excluded after spectroscopic observation \citep{Ostensen2010}. KIC 10536147 was listed as a white dwarf candidate \citep{MNJMK2012}, and was thus omitted from the study of MS B stars \citep{BBDDDC2015}. However, we can rule out the possibility based on its LAMOST spectrum and updated parameters.

\begin{figure}[htbp]
\epsscale{}
\plotone{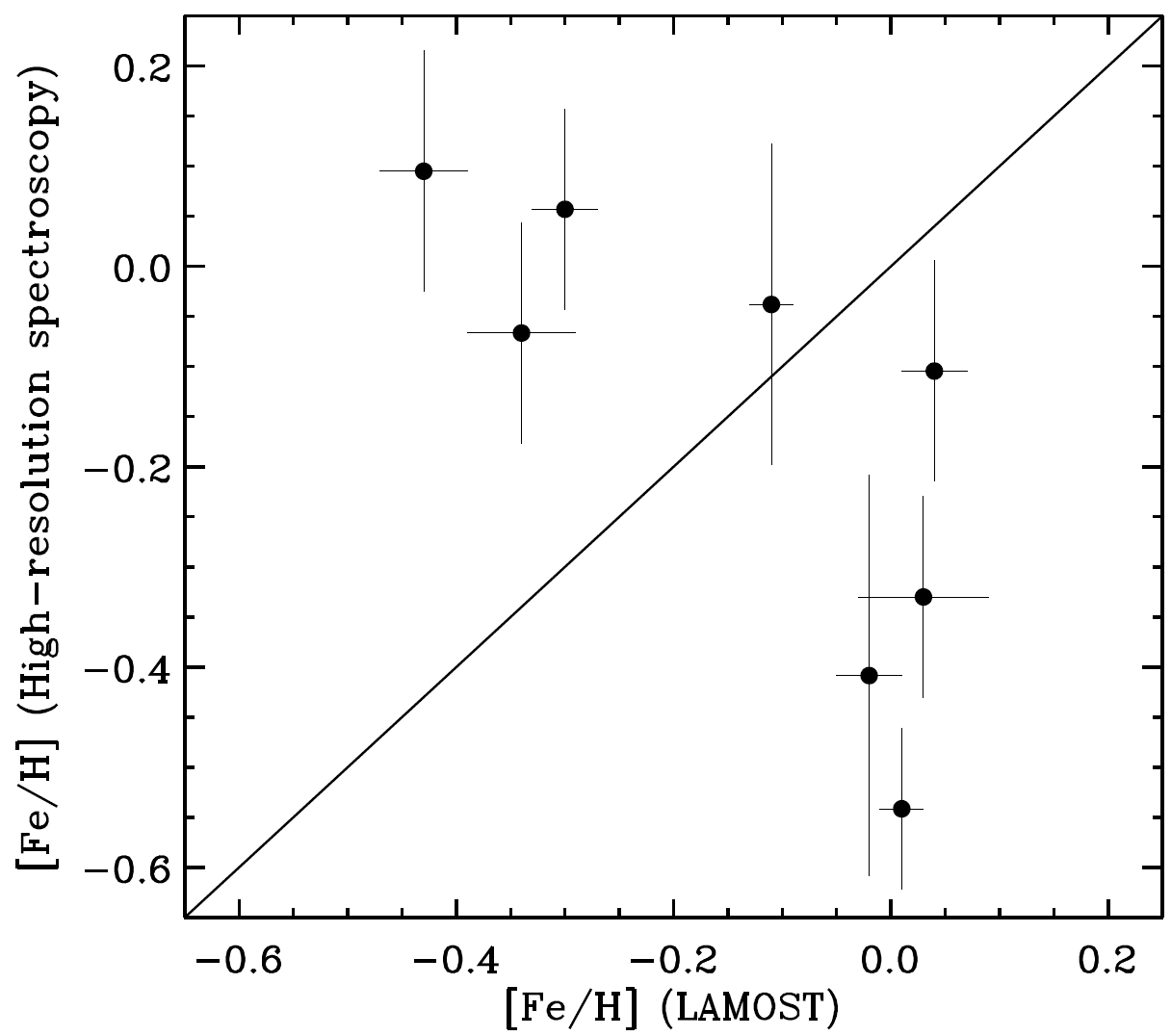}
\caption{Comparison of metallicity between LAMOST and high-resolution spectroscopy.}
\label{fig3}
\end{figure}

Only eight stars in Table~\ref{tab1} have metallicity derived from high-resolution spectroscopy. The comparison between the LAMOST metallicities and those derived from high-resolution spectroscopic data is shown in Figure~\ref{fig3}. The discrepancies are obvious, even when the external uncertainty in LAMOST [Fe/H] is taken into account. Unlike $T_\mathrm{eff}$ and $\log g$, the estimation of metallicity from low-resolution spectra of B stars is less reliable due to the lack of metallic lines. Nevertheless, the results shown in Figure~\ref{fig3} are surprising, where the LAMOST and spectroscopic values are almost anticorrelated. Further systematic comparisons based on larger samples are needed to assess the reliability of LAMOST metallicity of B stars.

\section{Identification of four B-type variable stars}\label{sec3}

As shown in the previous section, the underestimation in KIC $T_\mathrm{eff}$ and overestimation in KIC $\log g$ resulted in the misclassification of B stars. The consistency between LAMOST and high-resolution spectroscopy proves that we can use LAMOST data to identify such misclassified stars. We searched in LAMOST DR2 for B stars in the {\it Kepler} field with S/N$(g) \ge 50$ and obtained another four MS B stars with large $T_\mathrm{eff}$ discrepancies between the KIC and LAMOST ($\Delta T_\mathrm{eff}/T_\mathrm{eff}\mathrm{(LAMOST)}\simeq -50\%$). Their stellar parameters are listed in Table~\ref{tab2}, including the radial velocity, $v_\mathrm{r}$, derived from LAMOST spectra. The $T_\mathrm{eff}$ differences between the KIC and LAMOST are shown in Figure~\ref{fig1} as open circles. The four stars have very similar $T_\mathrm{eff}$ and $\log g$, and their metallicities are close to the solar value.

\begin{deluxetable*}{ccccccccr}
\tablewidth{0pt}
\tablecaption{Stellar parameters of the four B stars.\label{tab2}}
\tablehead{
\colhead{KIC Number} & \multicolumn{3}{c}{KIC} && \multicolumn{4}{c}{LAMOST} \\
& \colhead{$T_\mathrm{eff}$ (K)} & \colhead{$\log g$} & \colhead{[Fe/H]} && \colhead{$T_\mathrm{eff}$ (K)} & \colhead{$\log g$} & \colhead{[Fe/H]} & \colhead{$v_\mathrm{r}$ (km s$^{-1}$)}}
\startdata
5309849 & 9084 & 4.058 & \phn0.048 && 16369\,$\pm$\,185    & 3.72\,$\pm$\,0.12 &    -0.17\,$\pm$\,0.07 &  -12.1\,$\pm$\,6.6\\
6462033 & 8387 & 4.315 & \phn0.276 && 16808\,$\pm$\,93\phn & 3.73\,$\pm$\,0.06 &    -0.04\,$\pm$\,0.03 &    4.6\,$\pm$\,3.0\\
8255796 & 8583 & 3.971 & \phn0.023 && 16622\,$\pm$\,118    & 3.71\,$\pm$\,0.07 & \phn0.01\,$\pm$\,0.04 & -117.8\,$\pm$\,3.9\\
8324482 & 8159 & 3.819 &    -0.017 && 16469\,$\pm$\,71\phn & 3.72\,$\pm$\,0.05 &    -0.06\,$\pm$\,0.03 &  -18.5\,$\pm$\,2.4\\
\enddata
\tablecomments{The LAMOST errors are internal ones (see Section \ref{sec2} for realistic error estimates).}
\end{deluxetable*}

\begin{figure}[htbp]
\centering
\includegraphics[scale=0.4]{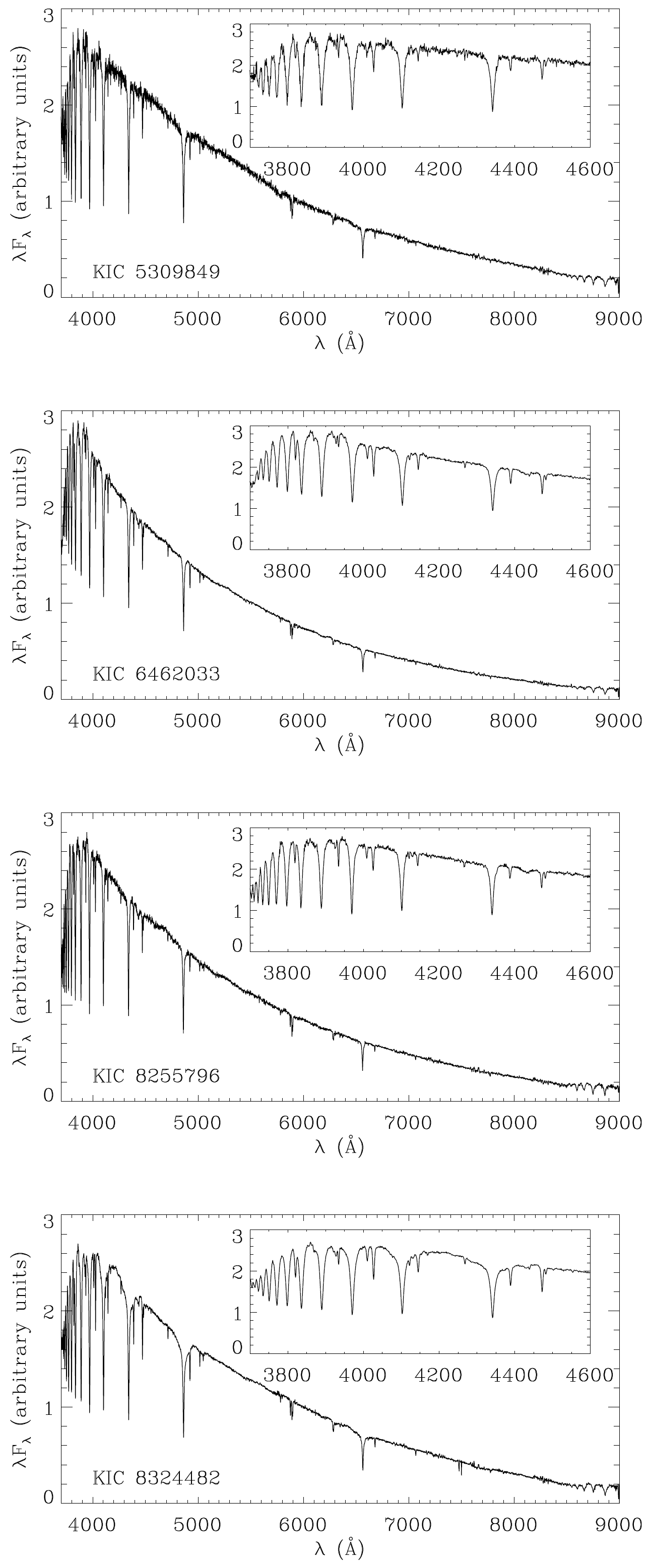}
\caption{LAMOST spectra of the four B stars. The inserts are close-up views of the blue ends.}
\label{fig4}
\end{figure}

Figure~\ref{fig4} shows the LAMOST spectra of these stars. According to \citet{GC2009}, we compare He I $\lambda4471$ and Mg II $\rm\lambda4481$ for the four apparent B-type spectra and find that they are earlier than $\sim$B5, consistent with the $T_\mathrm{eff}$ estimates. The relatively broad Balmer lines indicate that all of the samples are MS stars.

Among the four samples, KIC 8324482 shows special broad wings in H$\gamma$ and H$\delta$. This feature is very rare in normal B-type stars. However, although the Balmer lines show wide wings, the weak lines, e.g. He I and Mg II lines, do not show clear broadening, which is not fully understood.

\citet{USIU2014} carried out spectroscopic observations of KIC 6462033. By fitting H$\alpha$ and H$\beta$ line profiles, they found that the KIC $T_\mathrm{eff}$ was too high. Their best fit yielded $T_\mathrm{eff}=7150$ K. However, hydrogen line-profile fit does not necessarily lead to conclusive determinations, because hydrogen lines reach a maximum at $\sim$A2 on the MS and weaken toward both higher and lower temperatures. On the other hand, the presence of helium lines provides strong evidence of hot stars.

\begin{figure}[htbp]
\epsscale{}
\plotone{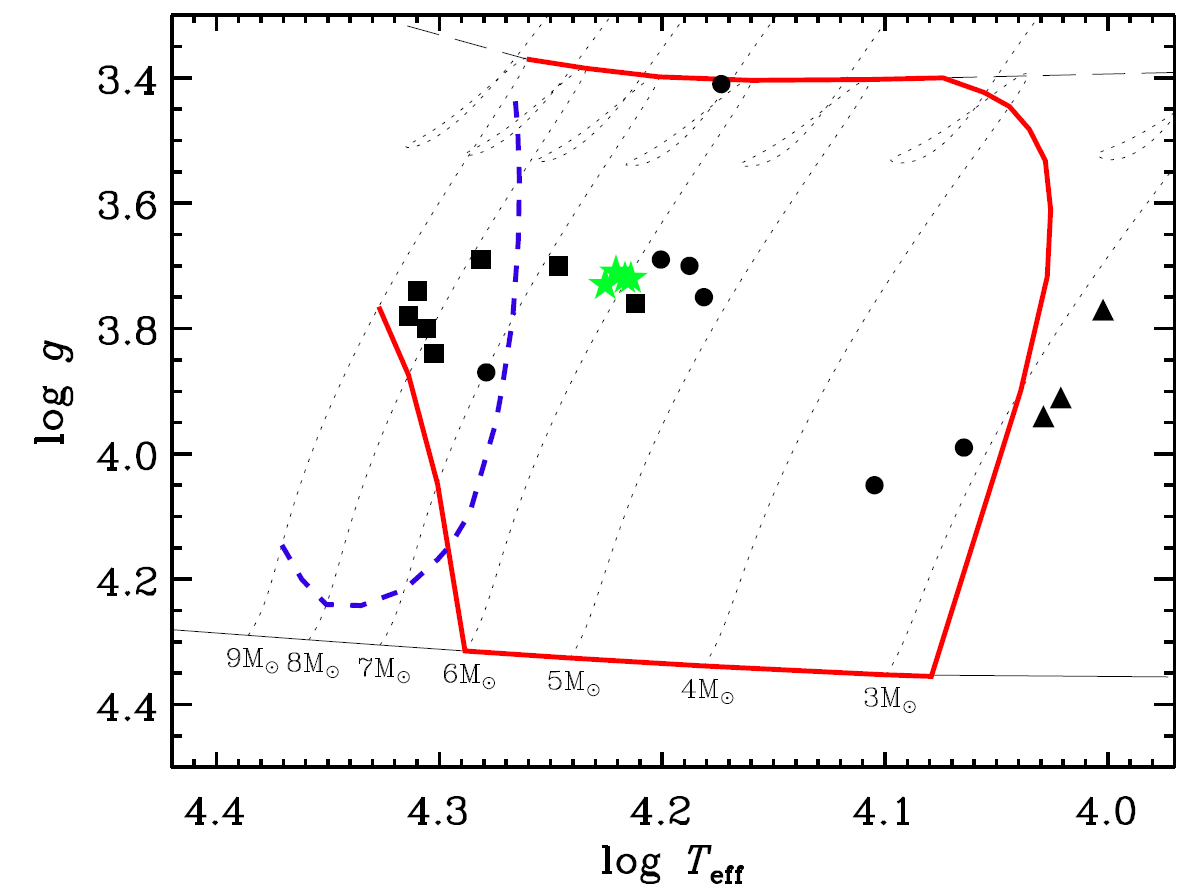}
\caption{Keil diagram of MS B stars using LAMOST stellar parameters. The green stars show the four misclassified B stars reported in this paper. The black symbols are the 17 B stars in Table~1 as classified by \citet{Balonaetal2011} and \citet{MNJMK2012}: circles are SPB stars, squares are hybrid stars, and triangles are stars with rotational modulation. The dotted lines show evolutionary tracks for selected masses computed with \texttt{MESA} using the same input physics as in Section~3.1 of \citet{MESA2015}. The solid and long-dashed black lines represent the zero-age main sequence and terminal-age mian sequence, respectively. The boundaries of the SPB (thick solid red line) and $\beta$ Cep (thick dashed blue line) instability strips are plotted for $M \le 9\mathrm{M}_\odot$ models (from \citealt{MESA2015}).}
\label{fig5}
\end{figure}

Using LAMOST $T_\mathrm{eff}$ and $\log g$, the four stars are placed in the Keil diagram in Figure~\ref{fig5} as green stars. The evolutionary tracks (dotted lines) were calculated using the \texttt{MESA} code \citep{MESA2015} with OPAL opacities \citep{OPAL} and proto-solar abundances from \citet{AGSS2009}. \citet{MESA2015} derived the theoretical instability strips of massive stars for oscillation modes of spherical degree $l=0$--$3$. The boundaries of the combined instability strips for $l \le 3$ modes were converted from the H-R diagram through stellar models and plotted as a solid red line for SPB stars and a dashed blue line for $\beta$ Cep stars in Figure~\ref{fig5}. The four stars are located near the $6M_\odot$ track and well within the SPB instability strip.

For comparison, the positions and oscillation types of 17 stars in Table~1 are shown as filled black symbols based on their LAMOST stellar parameters and the classification by \citet{Balonaetal2011} and \citet{MNJMK2012}. They are in good agreement with the theoretical instability strips. The three stars with rotational modulation are cooler than the red edge of the SPB instability strip. All of the other 14 stars are within the SPB instability strip, and indeed show low-frequency oscillations. The hybrid stars are generally hotter than pure SPB stars, and mostly within the overlapping region of the SPB and $\beta$ Cep instability strips. The oscillations of MS B stars are excited by the $\kappa$ mechanism operating in the partial ionization zone of iron-group elements, therefore the sizes of the theoretical instability strips and the overlapping region are sensitive to the adopted metallicity and the calculation of opacity in the Z-bump. Recent updates in the opacity calculations predict wider theoretical instability strips for MS B stars \citep{WFCKG2015, Moravveji2016}.

\section{Oscillation properties}\label{sec4}

The oscillation properties of all four stars have been previously studied. \citet{DBADR2011} carried out an automated variability analysis of all $\sim 150,000$ public {\it Kepler} light curves from the first quarter (Q1). KIC 5309849, KIC 6462033, KIC 8255796, and KIC 8324482 were assigned with the largest probability to the categories rotational modulation, active stars, miscellaneous, and $\gamma$ Dor stars, respectively. As noted by \citet{UMGetal2011}, the classifier used only three independent frequencies with the highest amplitudes, and no external information was taken into account. Based on KIC parameters and light curves with longer time coverage, KIC 6462033 \citep{UMGetal2011, Balona2014, USIU2014} and KIC 8324482 \citep{Balona2014} were identified as $\gamma$ Dor pulsators, while KIC 8255796 was classified as a rotating A star \citep{Balona2013}. As shown in Figure~\ref{fig5}, these four stars all fall into the instability strip of SPB stars with updated stellar parameters. We discuss their variability based on LAMOST parameters and new frequency analysis using all four years of the {\it Kepler} data.

\subsection{{\it Kepler} photometry}\label{sec4.1}

{\it Kepler} data are available in two cadences: long cadence (LC) with 29.4-min integrations and short cadence (SC) with 58.9-s integrations. LC data are delivered in quarters (Q0--Q17). Most quarters lasted for approximately one fourth of {\it Kepler}'s orbital period of 372.5\,d, which is the time interval between two satellite rolls performed to keep the solar panels pointing toward the Sun. Q0 (9.7\,d), Q1 (33.5\,d), and Q17 (31.8\,d) are short quarters. SC data are further divided into months, corresponding to the 32-d interval between two data downlinks. At any given time, only 512 targets were observed in SC mode. In our sample of four B stars, SC data are only available for KIC 6462033 for one month, while LC data are almost continuous throughout the mission for all four stars (Q1-Q17 for KIC 8255796 and Q0-Q17 for the other three). Therefore we use LC data in the following analysis. The time span $\Delta T$ is 1459.5\,d for Q1--Q17 data, and 1470.5\,d for Q0--Q17 data. The corresponding Rayleigh frequency resolution is $1/\Delta T \approx 0.0007$\,d$^{-1}$. The Nyquist frequency of LC data is 24.5\,d$^{-1}$.

Both light curve files and target pixel files can be downloaded from the Mikulski Archive for Space Telescopes\footnote[2]{http://archive.stsci.edu}. The light curves were extracted from the target pixel files using standard masks, which sometimes missed pixels with significant flux. Therefore we re-extracted the light curves from the target pixel files using custom masks. The custom masks were designed for each quarter separately by containing all of the pixels with significant flux and avoiding possible contamination. The extraction produces new simple aperture photometry (SAP) light curves. Systematic artifacts were then removed from the SAP light curves by subtracting the cotrending basis vectors (CBVs). The use of at least two CBVs was recommended. In practice, we found that 2--4 CBVs usually sufficed for artifact removal. We then filtered out all of the bad data points with SAP\_QUALITY $\ne 0$, and converted the fluxes to part per million (ppm) around the mean value for each quarter.

\begin{figure}[htbp]
\epsscale{}
\plotone{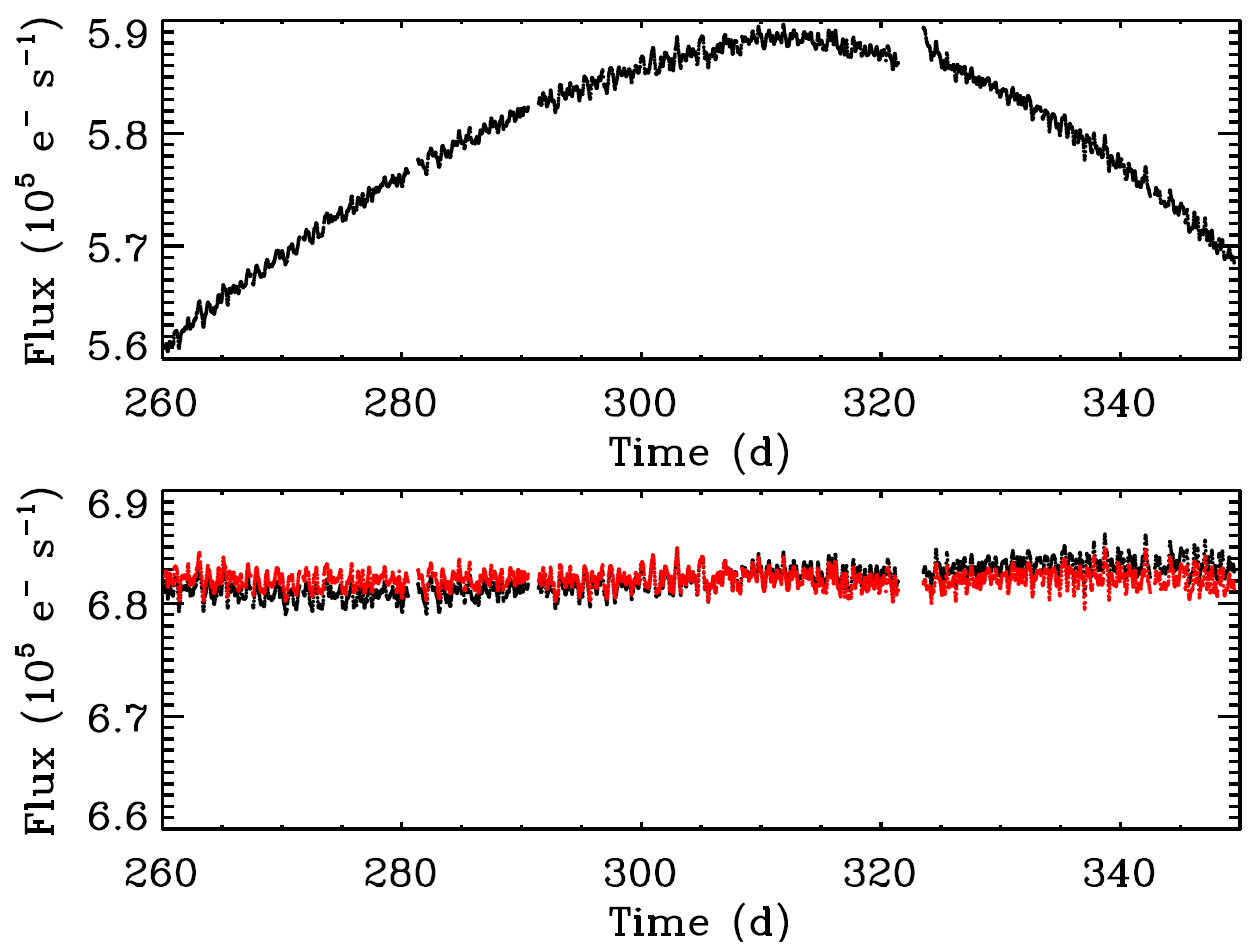}
\caption{Comparison of light curves of Q3 LC data of KIC 6462033. Top panel: SAP light curve using the standard mask. Bottom panel: new SAP light curve (black) using the custom mask and cotrended light curve (red) using 3 CBVs. Notice the different flux levels in the two panels.}
\label{fig6}
\end{figure}

The creation of custom masks, extraction, and cotrending of new light curves were carried out using the software \texttt{PyKE} \citep{PyKE}. Figure~\ref{fig6} shows an example of the improvement in extracted light curves. As can be seen, the new light curves show significantly less instrumental trends and larger variability amplitude.

\subsection{Frequency analysis}\label{sec4.2}

We used the software \texttt{SigSpec} \citep{SigSpec} to carry out the frequency analysis. It performs discrete Fourier transform and least-square fitting iteratively and automatically. This greatly facilitates the extraction of a large amount of frequencies, which is common in the analysis of space photometry. For a given time series, \texttt{SigSpec} carries out step-by-step prewhitening of the most significant signal components. When a new peak is extracted at each step, the software fits all the detected frequencies, amplitudes, and phases simultaneously. The error estimation of frequency, amplitude, and phase follows the formulas given by \citet{KRW2008}. If the frequency separation is less than three times the Rayleigh frequency resolution, the appropriate upper limit of the frequency error is $1/(4\Delta T) \approx 0.00017$\,d$^{-1}$.

\texttt{SigSpec} uses the parameter {\it sig} to define the amplitude spectral significance. In the analysis of ground-based data, a frequency is usually accepted as significant if its amplitude S/N $\ge 4.0$ (Breger et al. 1993). This is approximately equivalent to {\it sig} $\ge 5.46$ \citep{SigSpec}. However, the criterion of S/N $\ge 4.0$ may not be suitable in the analysis of space-based observations because it often results in a huge number of closely spaced low-amplitude frequencies, which may not be all physical \citep[e.g.][]{Papics2013}. The cumulative errors in the iterative prewhitening process influence subsequent frequency extraction, and give rise to spurious frequencies \citep{VanReeth2015}. Therefore the number of low-amplitude frequencies is likely overestimated using the S/N $\ge 4.0$ criterion. In practice, a higher or more stringent limit of S/N \citep[e.g.][]{Papics2013} or {\it sig} \citep[e.g.][]{Chapellier2011} was often used, in order to avoid false detection and reduce the number of frequencies to a manageable amount. In this work, we adopted the {\it sig} $=20$ criterion\citep{Chapellier2011}, and extracted $\sim\,500$ frequencies for each star.

The extracted frequencies were then checked for combinations (including harmonics). A frequency was accepted to be a combination if the difference between it and the combination of independent frequencies was less than the Rayleigh frequency resolution $1/\Delta T$. If two frequencies were separated by less than $1/\Delta T$, they were deemed as indistinguishable \citep{Papics2012}. The chance of finding spurious combinations, which are a result of mere mathematical coincidence, increases with the number of detected frequencies \citep{Papics2012}. Except for KIC 8255794, we only considered second or third order combinations formed by two independent frequencies \citep{Papics2013, Papics2014}.

High-order g modes of the same degree in SPB stars are expected to show equidistant period spacing in non-rotating homogeneous approximation \citep{Tassoul1980}. The chemical gradient near the core causes oscillatory deviations to the constant period spacing which can be further modified by extra mixing \citep{Miglio2008}, and stellar rotation introduces shifts in the period spacing series \citep{Bouabid2013}. Therefore the period spacing patterns in SPB stars are sensitive probes of stellar interior. After excluding true combinations, we searched for possible quasi-equidistant period spacings and possible distorted period series. The former can be detected in the autocorrelation function \citep{Papics2014}, while the latter requires manual identification \citep{Papics2017}.

\subsection{Results for individual stars}\label{sec4.3}

\begin{figure*}[htbp]
\centering
\gridline{\fig{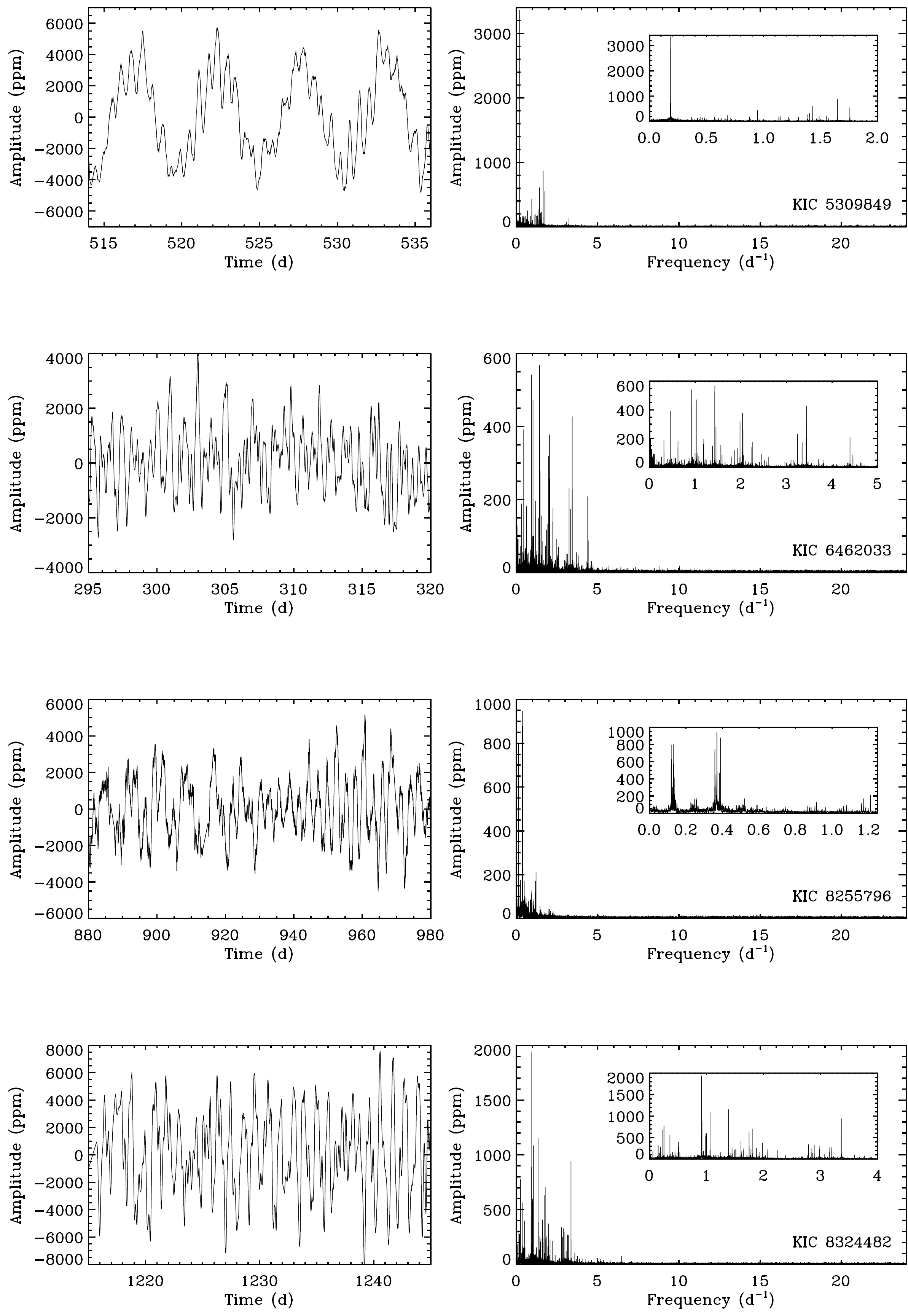}{0.8\textwidth}{}}
\caption{Light-curve segments (left panels) and amplitude spectra (right panels) of the four B stars. The inserts in the right panels are close-up views of low frequencies.}
\label{fig7}
\end{figure*}

Figure~\ref{fig7} shows light-curve segments and the amplitude spectra of the four stars using all four years of the data. Low frequencies are clearly seen in each panel. Although the stars are very close in the Keil diagram (Figure~\ref{fig5}), their amplitude spectra show different characteristics. We describe the results of the frequency analysis for each star.

\subsubsection{KIC 5309849}\label{sec4.3.1}

We extracted 512 significant frequencies with $sig > 20$. The light curve and amplitude spectrum of KIC 5309849 are dominated by the most significant frequency $f_1=0.18505$\,d$^{-1}$ with much higher amplitude and significance than the other frequencies. $f_1$ was detected as the rotation frequency \citep{NGSK2013, RRB2013}, and we do find its harmonic $2f_1$, which is usually taken as a sign of rotation \citep{Balona2013}. However, there is also the possibility of ellipsoidal variability, in which case KIC 5309849 is a close binary, and the very stable $f_1$ is related to orbital variation. Unfortunately, we do not have multi-epoch LAMOST observations to confirm its binary nature.

\begin{figure}[htbp]
\epsscale{}
\plotone{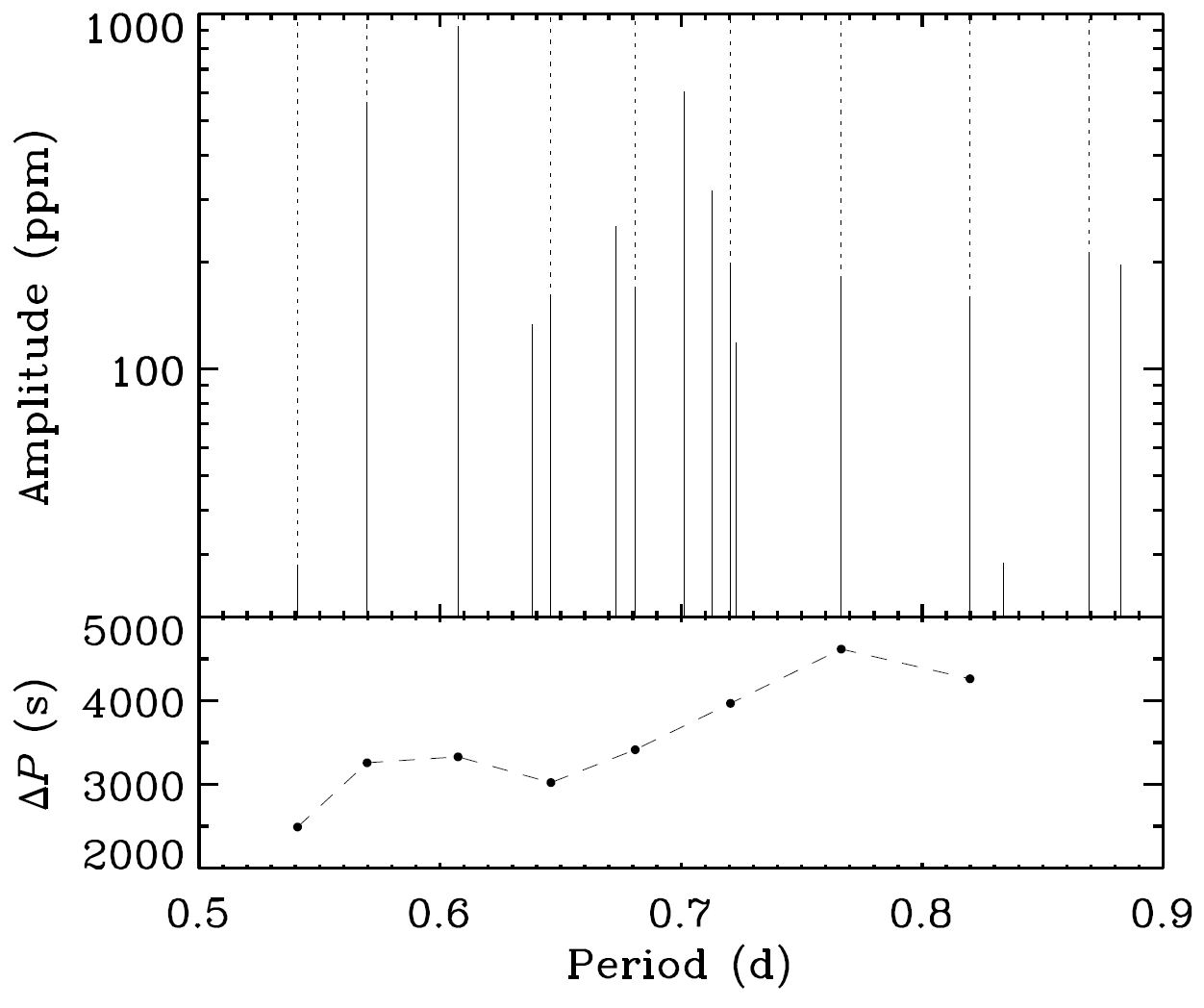}
\caption{Period spaing series in KIC 5309849. Top panel: a close-up view of the amplitude spectrum of KIC 5309849. The peaks of the period spacing series are marked with dashed lines. Bottom panel: observed period spacings. Error bars are smaller than the symbol size.}
\label{fig8}
\end{figure}

The classification of rotational modulation or ellipsoidal variable is incomplete to describe the variability of KIC 5309849. After extracting $f_1$, there are still many significant low frequencies below 2\,d$^{-1}$ that are most likely caused by pulsation. We managed to detect a short tilted period series consisting of nine consecutive radial orders among the high-amplitude frequencies, as listed in Table~\ref{tab:5309849} and shown in Figure~\ref{fig8}. The upward slope indicates that these modes are retrograde modes shifted by rotation \citep{VanReeth2015}. Considering the stellar parameters, the detected period spacing values are not likely between consecutive dipole modes, but of higher degree. The compatibility of $f_1$ being the rotation frequency and the observed period spacing needs further confirmation from theoretical modeling.

\begin{table}
\centering
\caption{Period spaing series in KIC 5309849.}\label{tab:5309849}
\begin{tabular}{ccccc}
\tablewidth{0pt}
\hline
\hline
& \colhead{Frequency}     & \colhead{Period} & \colhead{Amplitude} & \colhead{{\it sig}} \\
& \colhead{(d$^{-1}$)}    & \colhead{(d)}    & \colhead{(ppm)}     &                     \\
& \colhead{$\pm$\,0.00017}&                  &                     &                     \\
\hline
  1  &  1.15055  &  0.86915  &     209\,$\pm$\,8\phn   &   628  \\
  2  &  1.21977  &  0.81983  &     156\,$\pm$\,7\phn   &   450  \\
  3  &  1.30480  &  0.76640  &     178\,$\pm$\,7\phn   &   508  \\
  4  &  1.38798  &  0.72047  &     194\,$\pm$\,6\phn   &   856  \\
  5  &  1.46852  &  0.68096  &     166\,$\pm$\,12      &   177  \\
  6  &  1.54803  &  0.64598  &     158\,$\pm$\,6\phn   &   619  \\
  7  &  1.64623  &  0.60745  &     905\,$\pm$\,15      &  3417  \\
  8  &  1.75519  &  0.56974  &     552\,$\pm$\,11      &  2228  \\
  9  &  1.84875  &  0.54091  &  \phn27\,$\pm$\,3\phn   &    67  \\
\hline
\end{tabular}
\end{table}

Besides a large number of low frequencies, we have also extracted six high frequencies, as listed in Table~\ref{tab:hfreq}. They are well separated from the low frequencies, and cannot be explained as low-order combinations. The first three frequencies in Table~\ref{tab:hfreq} have been detected in each of the long quarters Q2--Q16. Thus they are stable over time, although their amplitudes are relatively low. Therefore, KIC 5309849 is a possible SPB/$\beta$ Cep hybrid. In Figure~\ref{fig5}, KIC 5309849 lies outside the plotted theoretical $\beta$ Cep instability strip. Even when we consider the new theoretical instability strips of $\beta$ Cep and SPB stars by \citet{Moravveji2016}, which were calculated based on the improved opacity tables, KIC 5309849 is still cooler then the red edge of the overlapping region where the hybrids are expected to occur.

\begin{table}
\centering
\caption{High frequencies of KIC 5309849.}\label{tab:hfreq}
\begin{tabular}{cccc}
\tablewidth{0pt}
\hline
\hline
& \colhead{Frequency}     & \colhead{Amplitude} & \colhead{{\it sig}} \\
& \colhead{(d$^{-1}$)}    & \colhead{(ppm)}     &                     \\
& \colhead{$\pm$\,0.00017}&                     &                     \\
\hline
  1  &  23.80189  &     32\,$\pm$\,4   &   79  \\
  2  &  23.67482  &     30\,$\pm$\,4   &   70  \\
  3  &  18.14375  &     15\,$\pm$\,3   &   36  \\
  4  &  23.26568  &     13\,$\pm$\,2   &   32  \\
  5  &  22.89368  &     11\,$\pm$\,2   &   27  \\
  6  &  23.38008  &     10\,$\pm$\,2   &   22  \\
\hline
\end{tabular}
\end{table}

\subsubsection{KIC 6462033}\label{sec4.3.2}

KIC 6462033 shows a rich spectrum of low frequencies. \citet{UMGetal2011} reported $46$ independent frequencies, whereas \citet{USIU2014} identified five independent frequencies and hundreds of combinations and harmonics. Our selection criterion resulted in 509 frequencies. We also checked the amplitude spectrum of SC data in Q3.3, and did not find significant peaks beyond 24.5\,d$^{-1}$.

\begin{figure}[htbp]
\epsscale{}
\plotone{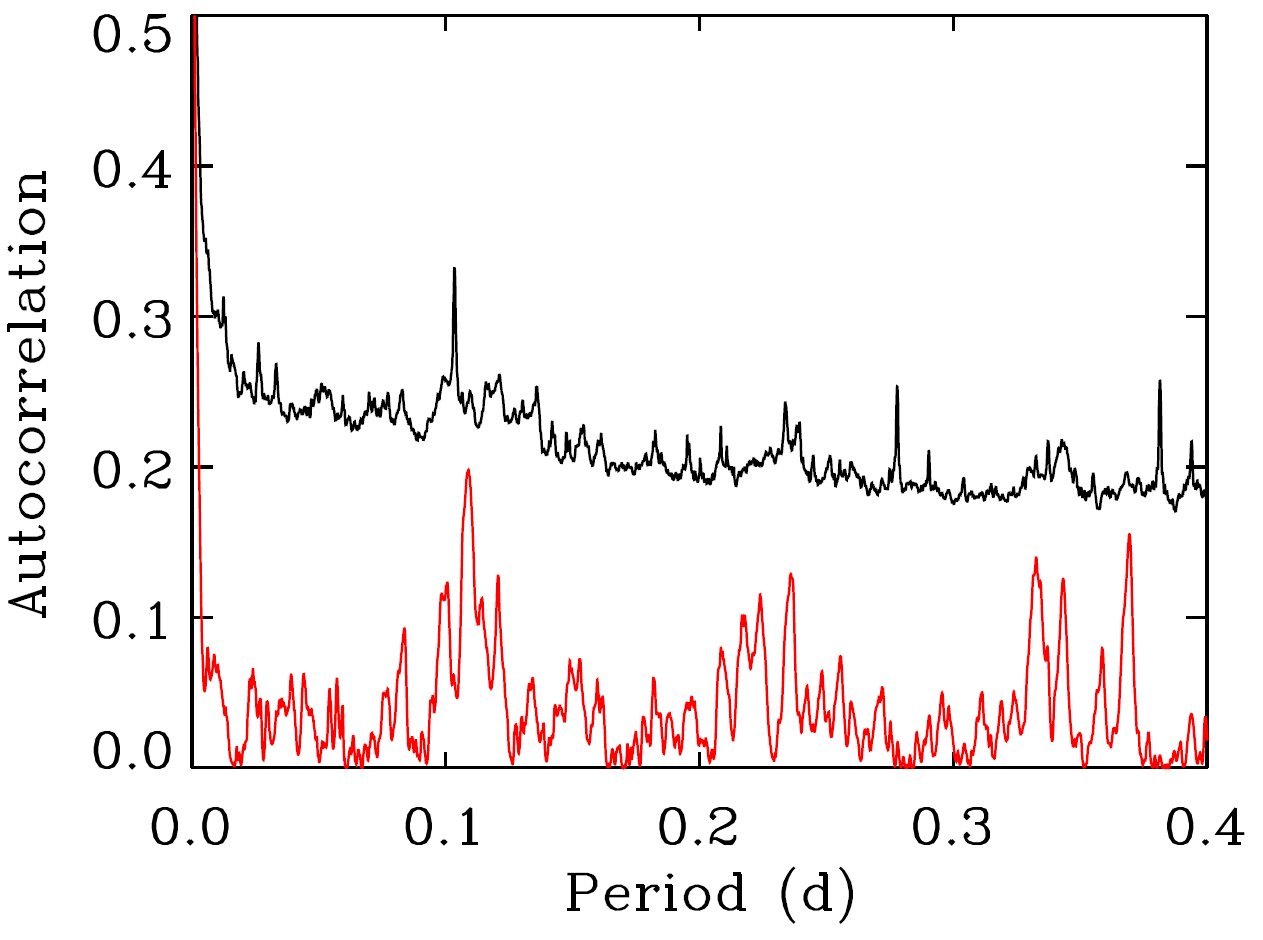}
\caption{Autocorrelation function of the original periodogram (black) and a modified periodogram where frequencies with $sig > 50$ were given an artificial power, while all others were given zero power (red), both calculated from the period range of 0.2\,d to 2.4\,d.}
\label{fig9}
\end{figure}

The frequencies of KIC 6462033 show a clear quasi-equidistant period spacing pattern in the period range of 0.2\,d to 2.4\,d. The black line in Figure~\ref{fig9} shows the autocorrelation function of the original periodogram (transformed into period space). Although the bumps of real period spacings are visible, the three highest peaks at 0.104\,d, 0.278\,d, and 0.382\,d correspond to the period differences between the three most significant frequencies. To avoid the strong influence of a few high-amplitude frequencies on the autocorrelation function, we gave the same artificial power to all of the frequencies with $sig >50$, and set the power of the other frequencies to zero. The autocorrelation function of the modified periodogram is shown in red in Figure~\ref{fig9}. We can easily identify the peaks at 0.109\,d and approximately twice and thrice this value.

\begin{figure}[htbp]
\epsscale{}
\plotone{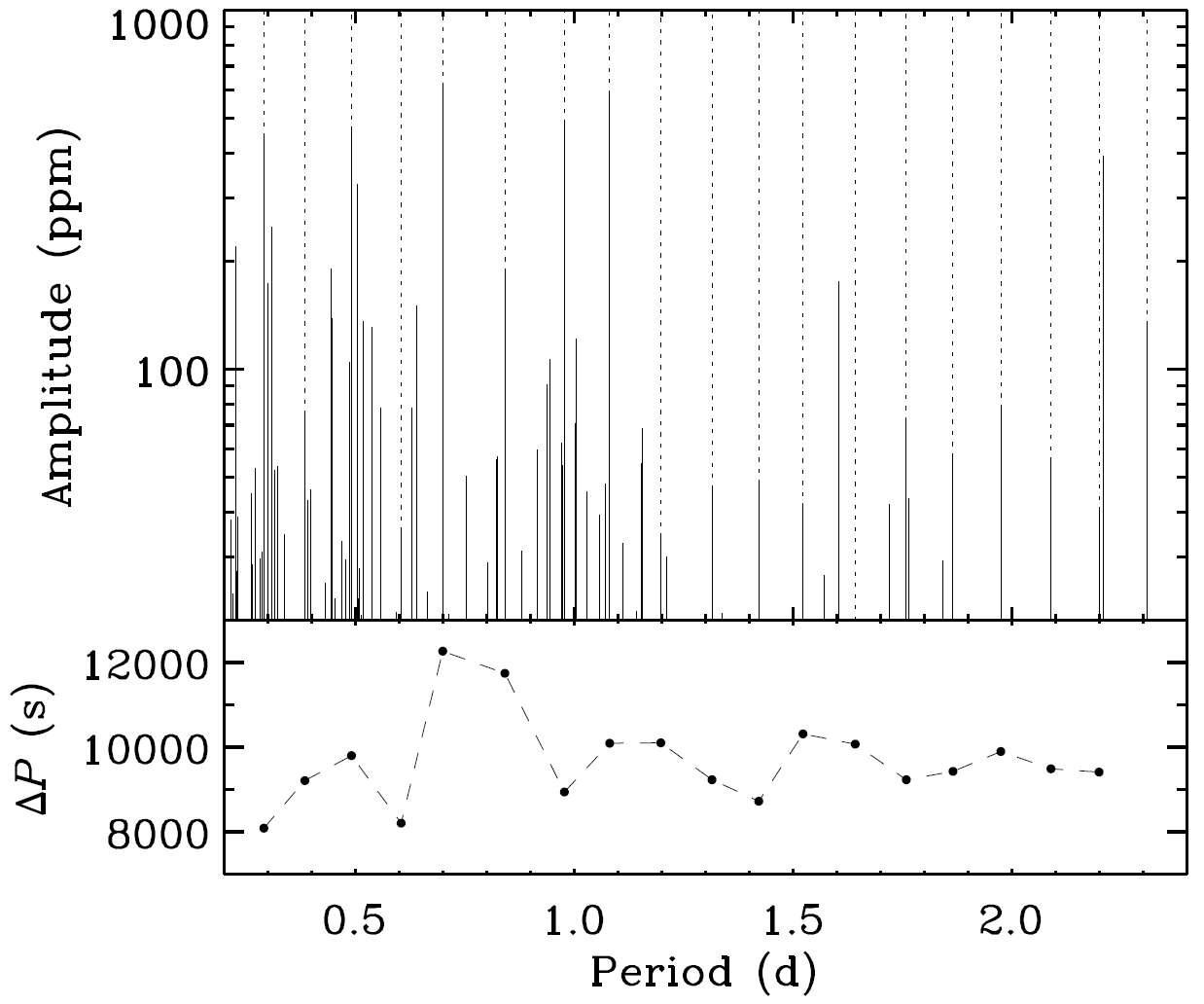}
\caption{The same as Figure~\ref{fig8}, but for KIC 6462033.}
\label{fig10}
\end{figure}

\begin{table}
\centering
\caption{Period spacing series in KIC 6462033.}\label{tab:6462033}
\begin{tabular}{ccccr}
\tablewidth{0pt}
\hline
\hline
& \colhead{Frequency}     & \colhead{Period} & \colhead{Amplitude} & \colhead{{\it sig}} \\
& \colhead{(d$^{-1}$)}    & \colhead{(d)}    & \colhead{(ppm)}     &                     \\
& \colhead{$\pm$\,0.00017}&                  &                     &                     \\
\hline
\phn1  & 0.43334  &  2.30764  &     132\,$\pm$\,12     &   108\\
\phn2  & 0.45481  &  2.19870  &  \phn40\,$\pm$\,4\phn  &    86\\
\phn3  & 0.47873  &  2.08887  &  \phn55\,$\pm$\,4\phn  &   125\\
\phn4  & 0.50651  &  1.97431  &  \phn77\,$\pm$\,9\phn  &    69\\
\phn5  & 0.53614  &  1.86518  &  \phn57\,$\pm$\,4\phn  &   137\\
\phn6  & 0.56872  &  1.75834  &  \phn71\,$\pm$\,5\phn  &   176\\
\phn7  & 0.60910  &  1.64177  &  \phn17\,$\pm$\,2\phn  &    44\\
\phn8  & 0.65686  &  1.52240  &  \phn41\,$\pm$\,4\phn  &    95\\
\phn9  & 0.70352  &  1.42143  &  \phn47\,$\pm$\,4\phn  &   110\\
   10  & 0.76069  &  1.31459  &  \phn46\,$\pm$\,4\phn  &   114\\
   11  & 0.83496  &  1.19766  &  \phn34\,$\pm$\,4\phn  &    69\\
   12  & 0.92521  &  1.08084  &     585\,$\pm$\,14     &  1677\\
   13  & 1.02317  &  0.97736  &     483\,$\pm$\,12     &  1481\\
   14  & 1.18847  &  0.84142  &     186\,$\pm$\,7\phn  &   614\\
   15  & 1.42972  &  0.69944  &     613\,$\pm$\,14     &  1778\\
   16  & 1.68451  &  0.59365  &  \phn20\,$\pm$\,3\phn  &    44\\
   17  & 2.03655  &  0.49103  &     462\,$\pm$\,12     &  1302\\
   18  & 2.60128  &  0.38443  &  \phn74\,$\pm$\,5\phn  &   207\\
   19  & 3.43849  &  0.29083  &     443\,$\pm$\,11     &  1400\\
\hline
\end{tabular}
\end{table}

We found a series of 19 consecutive periods with quasi-equidistant spacing, as listed in Table~\ref{tab:6462033} and marked in the top panel of Figure~\ref{fig10}. The period series includes the high-amplitude frequencies between 0.2\,d--1.1\,d and extends to longer period with lower amplitude. The mean period spacing is 9681\,s, compatible with the period spacing of dipole modes of a $\sim 6M_\odot$ MS star \citep{Miglio2008, Degroote2010}. The deviation from uniform period spacing is characterized by a periodic component, as shown in the bottom panel of Figure~\ref{fig10}. The period spacing pattern provides rich information about the evolutionary status and internal structure of the star \citep{Miglio2008}. The mean spacing and periodicity indicate that KIC 6462033 is in the middle stage of core hydrogen burning. The amplitude of the periodic component has a very large value ($\sim$ 12,000\,s) at $\sim$ 0.8\,d, and decreases as the period increases. The decreasing amplitude is a signature of extra mixing outside the convective core and a smooth gradient of chemical composition, otherwise we would see no amplitude variation with period. The detailed shape of the chemical composition gradient determines the amplitude of the periodic component \citep{Degroote2010}. However, it is also possible that the period 0.84142\,d is not a member of the period series, while the real peaks at $\sim$ 0.8\,d and $\sim$ 0.9\,d are not excited to detectable amplitudes. In this scenario, the period spacings would be much smaller at these peaks, and the mean period spacing of the period series would also be smaller.

Quasi-equidistant period spacings in MS B stars have been previously detected in CoRoT targets HD 50230 \citep{Degroote2010} and HD 43317 \citep{HD43317} with smaller numbers of consecutive dipole modes. So far the slow rotator KIC 10526294 has been the only one well-studied SPB star that shows quasi-equidistant period spacing \citep{Papics2014}. Its 19 consecutive dipole modes enables \citet{Moravveji2015} to put stringent constrains on core overshooting and diffusive mixing. \citet{Triana2015} have derived its internal rotation profile using the rotationally split multiplets. Rotationally affected period series have also been detected in SPB stars using space photometry \citep{Papics2015,Papics2017,Zwintz2017}, providing rich information about stellar internal structures \citep{MTAM2016}. With the detected long period spacing series, detailed seismic modeling of KIC 6462033 is also possible. However, we have not detected reliable rotational splitting in its periodogram.

\subsubsection{KIC 8255796}\label{sec4.3.3}

For KIC 8255796, we extracted 472 significant frequencies. The amplitude spectrum shows multiple frequency groups (FGs) with quasi-equidistant distribution. The most significant peaks are in the first and third groups at $\sim 0.12$\,d$^{-1}$ and $\sim 0.37$\,d$^{-1}$, respectively. An inspection of the quarterly amplitude spectra revealed unresolved FGs and amplitude variations. Table~\ref{tab:8255796} lists the 55 frequencies with {\it sig} $> 100$ using all four years of the data. The frequencies are arranged according to their FGs.

\begin{deluxetable*}{crrcccrrc}
\renewcommand\arraystretch{1.1}
\tabletypesize{\small}
\tablewidth{0pt}
\tablecaption{Significant frequencies of KIC 8255796.\label{tab:8255796}}
\tablehead{
\colhead{Frequency}  & \colhead{Amplitude} & \colhead{{\it sig}} & \colhead{Note} & & \colhead{Frequency}  & \colhead{Amplitude} & \colhead{{\it sig}} & \colhead{Note}\\
\colhead{(d$^{-1}$)} & \colhead{(ppm)}     &                     &                & & \colhead{(d$^{-1}$)} & \colhead{(ppm)}     & & \\
\colhead{$\pm$\,0.00017}&                  &                     &                & & \colhead{$\pm$\,0.00017}&                  & & }
\startdata
\multicolumn{4}{c}{FG0}                                   & & 0.36942 &1410\,$\pm$\,31    &2051 & $2f_1+f_2$ \\
0.02191 &  80\,$\pm$\,8\phn & 114 & $-f_1+f_2-f_3+f_4$    & & 0.37445 & 133\,$\pm$\,8\phn & 222 & $f_1+2f_2+f_3-f_4$ \\
\multicolumn{4}{c}{FG1}                                   & & 0.37702 & 119\,$\pm$\,8\phn & 183 & $2f_1-f_2+2f_3$ \\
0.11305 &  88\,$\pm$\,8\phn & 121 & $f_1-f_2+2f_3-f_4$    & & 0.38178 & 173\,$\pm$\,10    & 288 & $2f_1+f_4$ \\
0.11894 & 779\,$\pm$\,17    &2068 & $f_1$                 & & 0.38238 & 513\,$\pm$\,14    &1321 & $f_1+2f_2$ \\
0.12243 & 147\,$\pm$\,9\phn & 252 & $f_1-f_2+f_3$         & & 0.38856 & 338\,$\pm$\,18    & 330 & $2f_1-2f_2+2f_3+f_4$ \\
0.12554 & 210\,$\pm$\,11    & 330 & $f_1-2f_2+2f_3$       & & 0.38886 &1008\,$\pm$\,23    &1887 & $2f_1-2f_2+2f_3+f_4$ \\
0.12603 & 292\,$\pm$\,10    & 713 & $f_1-2f_2+2f_3$       & & 0.38947 & 118\,$\pm$\,11    & 110 & $f_1+2f_3$ \\
0.12881 & 378\,$\pm$\,12    & 968 & $2f_2-f_3$            & & 0.39560 & 101\,$\pm$\,8\phn & 127 & $3f_2$ \\
0.13198 & 762\,$\pm$\,17    &1885 & $f_2$                 & & 0.39762 &  92\,$\pm$\,7\phn & 144 & $4f_3-f_4$ \\
0.13530 & 378\,$\pm$\,12    & 950 & $f_3$                 & & \multicolumn{4}{c}{FG4}                 \\
0.13827 & 208\,$\pm$\,10    & 416 & $-f_2+2f_3$           & & 0.47690 &  81\,$\pm$\,7\phn & 112 & $3f_1+2f_2-f_4$ \\
0.14133 & 125\,$\pm$\,9\phn & 191 & $f_2-f_3+f_4$         & & 0.48852 &  86\,$\pm$\,7\phn & 118 & $3f_1+f_2$ \\
0.14424 & 137\,$\pm$\,9\phn & 219 & $f_4$                 & & 0.50775 & 101\,$\pm$\,8\phn & 149 & $f_1+5f_2-2f_3$ \\
\multicolumn{4}{c}{FG2}                                   & & 0.51452 & 108\,$\pm$\,8\phn & 167 & $f_1+3f_2$ \\
0.22606 & 114\,$\pm$\,8\phn & 182 & $2f_1+f_2-f_4$        & & 0.52086 & 146\,$\pm$\,9\phn & 258 & $f_1+f_2+2f_3$ \\
0.23551 & 100\,$\pm$\,8\phn & 144 & $f_1+3f_2-f_3-f_4$    & & \multicolumn{4}{c}{FG5}                 \\
0.23748 & 105\,$\pm$\,8\phn & 144 & $2f_1$                & & 0.58714 &  87\,$\pm$\,8\phn & 121 & $4f_1+2f_2+f_3-2f_4$\\
0.24055 &  76\,$\pm$\,7\phn & 104 & $2f_1-f_2+f_3$        & & 0.59291 &  90\,$\pm$\,8\phn & 139 & $4f_1+3f_3-2f_4$\\
0.24229 &  93\,$\pm$\,8\phn & 130 & $f_1+f_2+f_3-f_4$     & & \multicolumn{4}{c}{FG6}                 \\
0.24529 & 184\,$\pm$\,9\phn & 350 & $f_1+2f_3-f_4$        & & 0.74270 &  86\,$\pm$\,8\phn & 121 & $4f_1+f_2+f_3$\\
0.25440 & 147\,$\pm$\,9\phn & 222 & $f_1+f_3$             & & \multicolumn{4}{c}{FG7}                 \\
0.25572 & 127\,$\pm$\,8\phn & 248 & $2f_2+f_3-f_4$        & & 0.86648 &  82\,$\pm$\,8\phn & 115 & $4f_1+2f_2+2f_3-f_4$\\
0.25692 & 111\,$\pm$\,8\phn & 168 & $f_1-f_2+2f_3$        & & 0.90579 &  83\,$\pm$\,8\phn & 115 & $f_1+5f_2+2f_3-f_4$\\
0.26371 &  84\,$\pm$\,7\phn & 119 & $2f_2$                & & 0.91529 & 133\,$\pm$\,9\phn & 222 & $f_1+2f_2+5f_3-f_4$\\
\multicolumn{4}{c}{FG3}                                   & & \multicolumn{4}{c}{FG8}                 \\
0.34734 &  97\,$\pm$\,8\phn & 146 & $3f_1+f_3-f_4$        & & 1.07887 &  87\,$\pm$\,8\phn & 127 & $-f_1+6f_2+3f_3$\\
0.35253 & 141\,$\pm$\,8\phn & 249 & $4f_1-f_2-f_3+f_4$    & & \multicolumn{4}{c}{FG9}                 \\
0.35771 & 692\,$\pm$\,34    & 401 & $2f_1+2f_2-f_4$       & & 1.16084 & 121\,$\pm$\,9\phn & 192 & $3f_1+5f_2+f_4$\\
0.35788 &1220\,$\pm$\,24    &2449 & $2f_1+2f_2-f_4$       & & 1.17314 & 174\,$\pm$\,10    & 308 & $f_1+6f_2+3f_3-f_4$\\
0.36348 & 495\,$\pm$\,37    & 171 & $3f_1-2f_2+2f_3$      & & 1.18445 &  76\,$\pm$\,8\phn & 101 & $f_1+5f_2+3f_3$\\
0.36367 & 832\,$\pm$\,23    &1287 & $2f_1+2f_3-f_4$       & & 1.21078 & 208\,$\pm$\,10    & 397 & $7f_1+3f_2+2f_3-2f_4$\\
0.36923 & 790\,$\pm$\,28    & 751 & $2f_1+f_2$            & &         &                   &     &\\
\enddata
\end{deluxetable*}

B stars that show FGs in their periodograms were classified as a separate type of variation by \citet{Balonaetal2011}. The origin of FGs was suspected to be related to rotation due to the resemblance to Be stars. \citet{Kurtzetal2015} demonstrated that the FGs in $\gamma$ Dor, SPB, and Be stars could be explained as a few base frequencies and their combination frequencies (including harmonics). They interpreted the base frequencies as g-mode oscillations, and showed that ``the combination frequencies can have amplitudes greater than the base frequency amplitudes." For KIC 8255796, we have found that all the frequencies in Table~\ref{tab:8255796} can be explained by using only four independent base frequencies in FG1 and their combinations, although high-order combinations are needed to explain high FGs.

Amplitude modulation adds complexity to the amplitude spectrum of KIC 8255796. As can be seen in Table~\ref{tab:8255796}, there are several pairs of indistinguishable frequencies in FG1 and FG3 with frequency separations smaller than the Rayleigh frequency resolution, e.g. 0.36923\,d$^{-1}$ and 0.36942\,d$^{-1}$. These frequency pairs were derived from unresolved peaks in the amplitude spectrum as a result of amplitude modulation \citep{BK2014}. A single frequency value was unable to completely remove the signal from the amplitude spectrum because the value and amplitude of a frequency changed over four years, and caused an unresolved peak when all data were used.

\begin{figure}[htbp]
\epsscale{}
\plotone{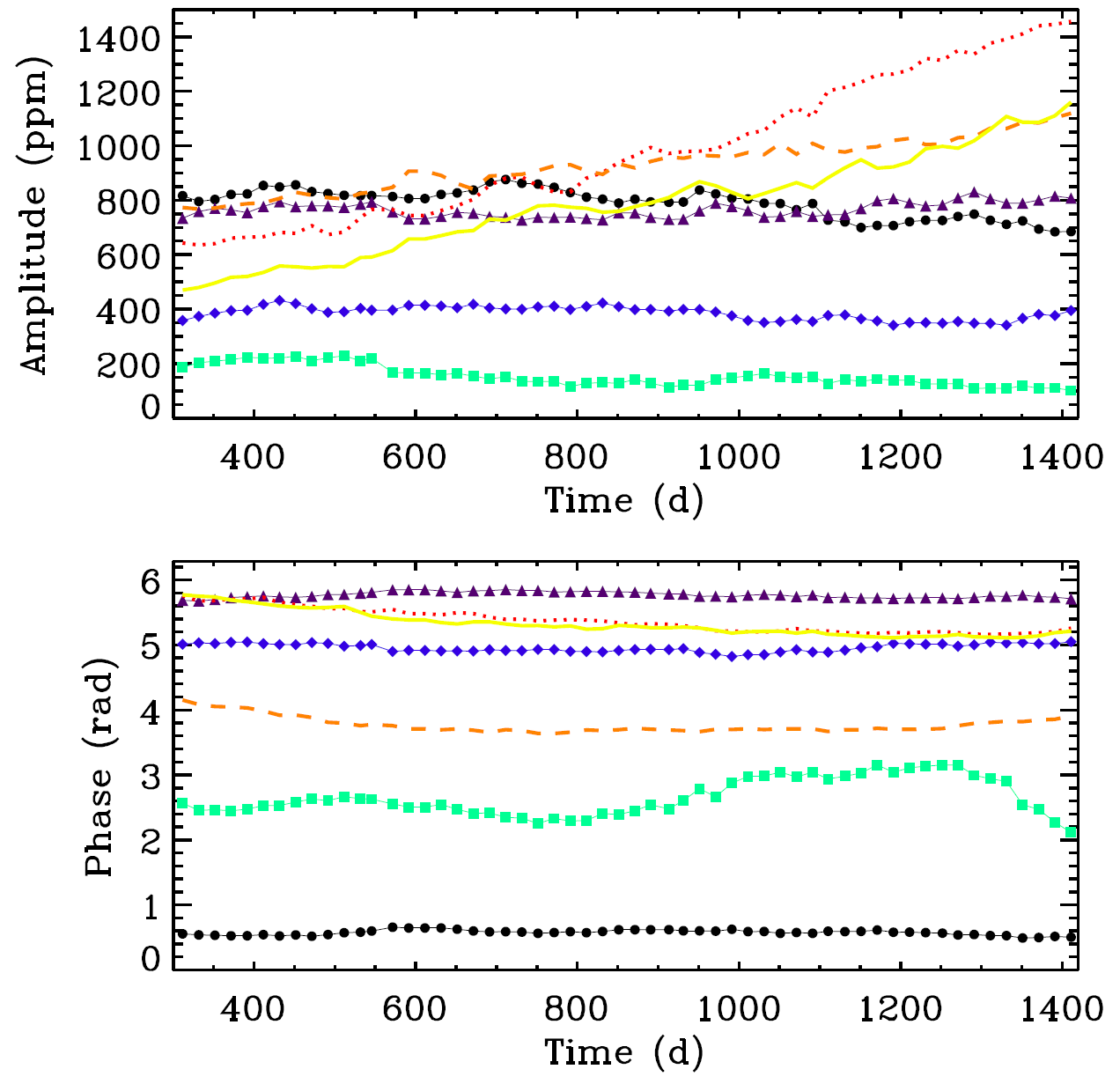}
\caption{Amplitude (top panel) and phase (bottom panel) variations of the four base frequencies in FG1 and three most significant frequencies in FG3: 0.11894\,d$^{-1}$ (black dots), 0.13198\,d$^{-1}$ (purple triangles), 0.13530\,d$^{-1}$ (blue diamonds), 0.14424\,d$^{-1}$ (green squares), 0.36942\,d$^{-1}$ (red dotted line), 0.38886\,d$^{-1}$ (orange dashed line), and 0.35788\,d$^{-1}$ (yellow solid line).}
\label{fig11}
\end{figure}

In order to track the variations, we carried out time-resolved analysis by calculating the amplitude spectra of 360-d subsets of the data moved in 20-day steps. For each subset, the amplitudes and phases were calculated with the frequencies fixed at the values in Table~\ref{tab:8255796} using the software \texttt{Period04} \citep{Period04}. Figure~\ref{fig11} shows the amplitude and phase variations of the four base frequencies in FG1, i.e. $f_1=0.11894$\,d$^{-1}$ (black dots), $f_2=0.13198$\,d$^{-1}$ (purple triangles), $f_3=0.13530$\,d$^{-1}$ (blue diamonds), and $f_4=0.14424$\,d$^{-1}$ (green squares), as well as the three most significant frequencies in FG3, i.e. 0.36942\,d$^{-1}$ (red dotted line), 0.38886\,d$^{-1}$ (orange dashed line), and 0.35788\,d$^{-1}$ (yellow solid line). The differences between FG1 and FG3 are evident. Except for $f_4$, which has a relatively low amplitude, the base frequencies have quite stable amplitudes and phases, while the frequencies in FG3 show significant and almost continuous amplitude growth over 4 yr. The amplitudes of 0.36942\,d$^{-1}$ and 0.35788\,d$^{-1}$ have increased by more than $100\%$, and their phases show very similar decreases. The significant frequencies in other FGs also show amplitude variations, but at lower amplitude levels. The coexistence of stable frequencies with frequencies having large amplitude variations may indicate a more complex origin of the FGs than combination effects, because one expects to see related variations of the base frequencies and their combinations, and the combination frequencies usually have much lower amplitudes than the base frequencies \citep{BKBMH2016}.

Amplitude modulation of p modes has been found in {\it Kepler} A stars \citep[e.g.][]{BK2014}. The ensemble study by \citet{BKBMH2016} shows that amplitude modulation is common among $\delta$ Sct stars. Various mechanisms can cause amplitude and/or phase variations, but the origin of variations in some stars remains unclear. Amplitude modulation has not been previously reported for B stars with FGs. A similar ensemble study would reveal whether such a phenomenon is common among these stars, and help the theoretical interpretation of FGs.

\subsubsection{KIC 8324482}\label{sec4.3.4}

\begin{figure}[htbp]
\epsscale{}
\plotone{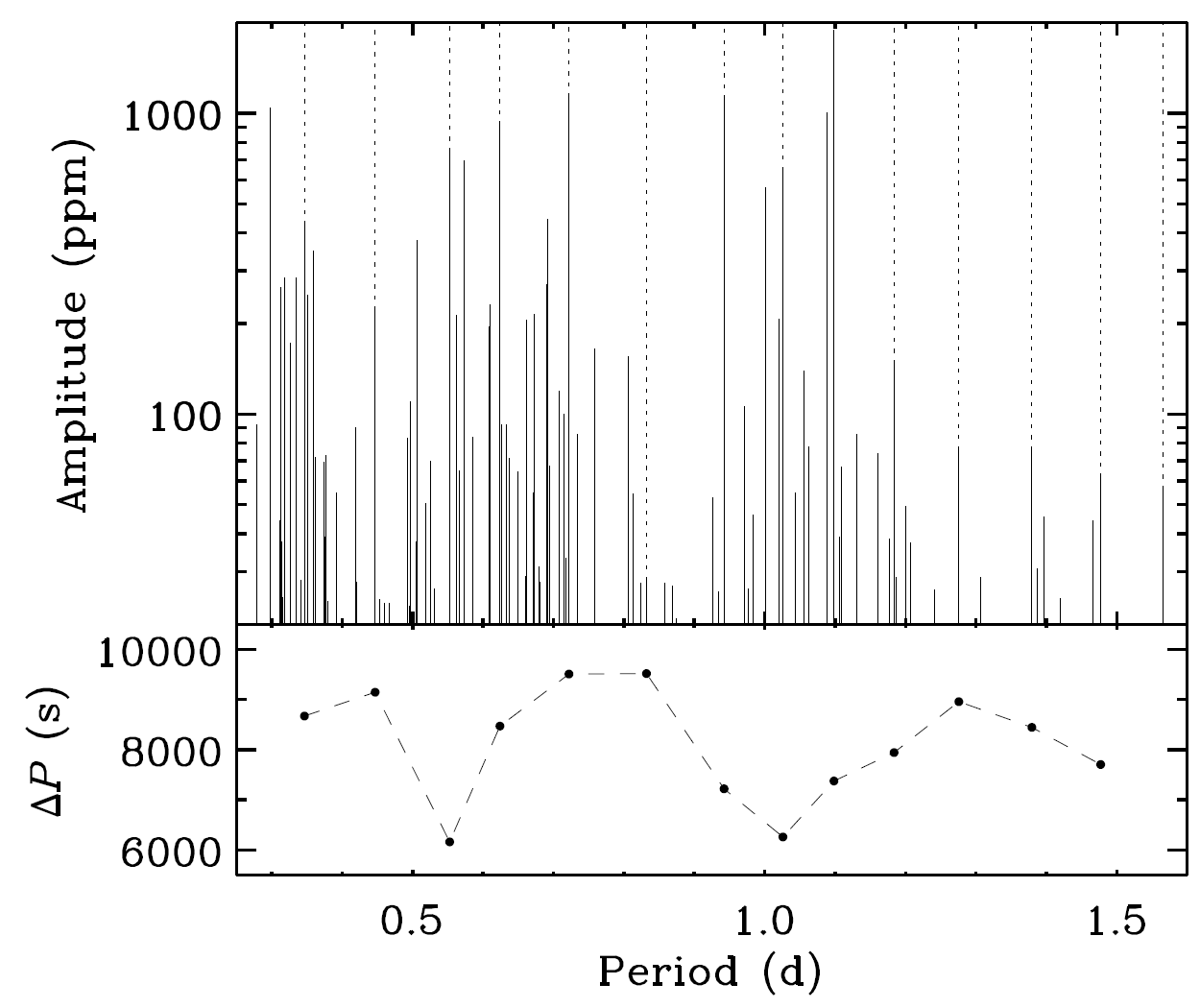}
\caption{The same as Figure~\ref{fig8}, but for KIC 8324482.}
\label{fig12}
\end{figure}

We extracted 446 frequencies for KIC 8324482. The dominant frequencies are mostly within the period range of 0.3\,d--1.1\,d. Due to the dense frequency spectrum, we were unable to detect clear period spacing using the modified autocorrelation function. We manually searched for possible period spacing patterns among high-amplitude frequencies, and managed to find a period series consisting of 14 consecutive dipole modes that show quasi-equidistant period spacing. The frequencies are listed in Table~\ref{tab:8324482} and marked in the top panel of Figure~\ref{fig12}.

The average spacing of the detected period series is 8108\,s, indicating a lower effective temperature compared with KIC 6462033. Again, we see periodicity in the observed period spacings, as shown in the bottom panel of Figure~\ref{fig12}. The period series is not long enough to draw conclusions on the amplitude variation of the periodic component, but the longer period suggests that KIC 8324482 is at an earlier evolutionary stage than KIC 6462033.

Our search for period spacing started with the most significant peaks in the periodogram \citep{Papics2017}. Less significant frequencies were then added after identifying possible period spacing. Therefore, dominant frequencies are selected with priority in this approach. It is possible that real peaks are ignored due to low significance, as discussed in Section \ref{sec4.3.2}.

\begin{table}
\centering
\caption{Period spacing series in KIC 8324482.}\label{tab:8324482}
\begin{tabular}{cccrr}
\tablewidth{0pt}
\hline
\hline
& \colhead{Frequency}     & \colhead{Period} & \colhead{Amplitude} & \colhead{{\it sig}} \\
& \colhead{(d$^{-1}$)}    & \colhead{(d)}    & \colhead{(ppm)}     &                     \\
& \colhead{$\pm$\,0.00017}&                  &                     &                     \\
\hline
\phn1  &  2.88685  &  0.34640  &   426\,$\pm$\,18     &   559\\
\phn2  &  2.23827  &  0.44677  &   221\,$\pm$\,11     &   347\\
\phn3  &  1.80946  &  0.55265  &   747\,$\pm$\,21     &  1206\\
\phn4  &  1.60262  &  0.62398  &   914\,$\pm$\,34     &   686\\
\phn5  &  1.38497  &  0.72204  &  1131\,$\pm$\,28     &  1553\\
\phn6  &  1.20177  &  0.83211  &    28\,$\pm$\,5\phn  &    22\\
\phn7  &  1.06125  &  0.94229  &  1116\,$\pm$\,28     &  1550\\
\phn8  &  0.97479  &  1.02587  &   646\,$\pm$\,19     &  1075\\
\phn9  &  0.91047  &  1.09833  &  1849\,$\pm$\,31     &  3433\\
   10  &  0.84479  &  1.18372  &   147\,$\pm$\,10     &   209\\
   11  &  0.78391  &  1.27566  &    76\,$\pm$\,7\phn  &    99\\
   12  &  0.72497  &  1.37936  &    76\,$\pm$\,8\phn  &    86\\
   13  &  0.67699  &  1.47712  &    61\,$\pm$\,8\phn  &    47\\
   14  &  0.63845  &  1.56631  &    56\,$\pm$\,7\phn  &    56\\
\hline
\end{tabular}
\end{table}

The use of KIC parameters lead to some confusion in the oscillation properties of KIC 6462033 and KIC 8324482. They were among the 12 high-temperature $\gamma$ Dor stars listed by \citet{Balona2014}. It was difficult to explain the existence of pure g-mode oscillators at $T_\mathrm{eff} \sim 8000$\,K. \citet{Balona2014} suggested the possibility of binary or incorrect $T_\mathrm{eff}$. We confirm here the latter to be the case for these two stars.

\section{Conclusion}\label{sec5}
Using LAMOST data, we have confirmed the underestimation in KIC $T_\mathrm{eff}$ and overestimation in KIC $\log g$ of B stars. The consistency of stellar parameters between LAMOST and other follow-up observations demonstrates the potential of LAMOST spectra in the variability classification and statistical study of early-type stars in the {\it Kepler} field.

Using LAMOST estimates of stellar parameters, our search for misclassifications led to the finding of four MS B stars which had been previously cataloged as A stars. We examined the spectra and confirmed their identities as MS B stars. Starting from {\it Kepler} target pixel files, we carried out detailed frequency analysis of the four stars. Although they lie very closely to each other in the Keil diagram, their amplitude spectra show very different characters. Based on the results of our analysis, we classify KIC 6462033 and KIC 8324482 as SPB stars, and KIC 5309849 as a candidate of SPB/$\beta$ Cep hybrid with rotational or ellipsoidal variability. The amplitude spectrum of KIC 8255796 shows both FGs and amplitude modulation. The cause of amplitude modulation and its relation to FGs require further investigation.

Our search for period spacing patterns lead to the detection of period series consisting of consecutive g modes in KIC 5309849, KIC 6462033, and KIC 8324482. The short tilted period series in KIC 5309849 indicates internal rotation, while the quasi-equidistant period spacing in KIC 6462033 and KIC 8324482 is a sign of slow rotation. The long period series in KIC 6462033 provides the opportunity for further seismic modeling and detailed studies of its near-core structure and internal mixing.

Our results prove the necessity of accurate stellar parameters in the study of stellar variability in the {\it Kepler} field. The combination of asteroseismology and ground-based follow-up observations will help one obtain more reliable estimations of stellar parameters \citep{DZZetal2014, Liuetal2015} and fully exploit the potential of {\it Kepler}'s high-precision photometry.

\acknowledgments
We thank the anonymous referee for helpful comments and suggestions that greatly improved the paper. This work was supported by National Natural Science Foundation of China (NSFC) through grants 11403039, 11633005, 11373032, 11333003, and 11403056, and the Joint Fund of Astronomy of NSFC and Chinese Academy of Sciences (CAS) through grants U1231119 and U1231202. C.Z. acknowledges support from the Young Researcher Grant of National Astronomical Observatories, CAS (NAOC). C.L. acknowledges support from the Strategic Priority Research Program ``The Emergence of Cosmological Structures" of CAS (Grant No. XDB09000000). J.F. acknowledges support from the National Basic Research Program of China (973 Program 2014CB845700 and 2013CB834900). Guoshoujing Telescope (the Large Sky Area Multi-Object Fiber Spectroscopic Telescope LAMOST) is a National Major Scientific Project built by CAS. Funding for the project has been provided by the National Development and Reform Commission. LAMOST is operated and managed by NAOC.



\begin{thebibliography}{99}
\bibitem[Asplund et al.(2009)]{AGSS2009}
Asplund, M., Grevesse, N., Sauval, A. J., \& Scott, P. 2009, ARA\&A, 47, 481
\bibitem[Balona(1994)]{Balona1994}
Balona, L. A. 1994, MNRAS, 268, 119
\bibitem[Balona(2013)]{Balona2013}
Balona, L. A. 2013, MNRAS, 431, 2240
\bibitem[Balona(2014)]{Balona2014}
Balona, L. A. 2014, MNRAS, 437, 1476
\bibitem[Balona \& Dziembowski(2011)]{BD2011}
Balona, L. A., \& Dziembowski, W. A. 2011, MNRAS, 417, 591
\bibitem[Balona et al.(2015)]{BBDDDC2015}
Balona, L. A., Baran, A. S., Daszy\'nska-Daszkiewicz, J., \& De Cat, P. 2015, MNRAS, 451, 1445
\bibitem[Balona et al.(2011)]{Balonaetal2011}
Balona, L. A., Pigulski, A., De Cat, P., et al. 2011, MNRAS, 413, 2403
\bibitem[Bohlender \& Monin(2011)]{BM2011}
Bohlender, D. A., \& Monin, D. 2011, AJ, 141, 169
\bibitem[Borucki et al.(2010)]{Kepler}
Borucki, W. J., Koch, D., Basri, G., et al. 2010, Science, 327, 977
\bibitem[Bouabid et al. (2013)]{Bouabid2013}
Bouabid M.-P., Dupret M.-A., Salmon S., et al. 2013, MNRAS, 429, 2500
\bibitem[Bowman \& Kurtz(2014)]{BK2014}
Bowman, D. M., \& Kurtz, D. W. 2014, MNRAS, 444, 1909
\bibitem[Bowman et al.(2016)]{BKBMH2016}
Bowman, D. M., Kurtz, D. W., Breger, M., Murphy, S. J., \& Holdsworth, D. L. 2016, MNRAS, 460, 1970
\bibitem[Breger et al.(1993)]{Breger1993}
Breger, M., Stich, J., Garrido, R., et al. 1993, A\&A, 271, 482
\bibitem[Brown et al.(2011)]{KIC}
Brown, T., Latham, D. W., Everett, M. E., \& Esquerdo, G. A. 2011, AJ, 142, 112
\bibitem[Chapellier et al.(2011)]{Chapellier2011}
Chapellier, E., Rodr\'iguez, E., Auvergne, M., et al. 2011, A\&A, 525, A23
\bibitem[Cui et al.(2012)]{LAMOST}
Cui, X., Zhao, Y., Chu, Y., et al. 2012, RAA, 12, 1197
\bibitem[Debosscher et al.(2011)]{DBADR2011}
Debosscher, J., Blomme, J., Aerts, C., \& De Ridder, J. 2011, A\&A, 529, A89
\bibitem[Degroote et al.(2010)]{Degroote2010}
Degroote. P., Aerts, C., Baglin, A., et al. 2010, Nature,464, 259
\bibitem[De Cat et al.(2015)]{DeCat2015}
De Cat, P., Fu, J., Ren, A., et al. 2015, ApJS, 220, 19
\bibitem[De Ridder et al.(2009)]{DeRidder2009}
De Ridder, J., Barban, C., Baudin, F., et al. 2009, Nature, 459, 398
\bibitem[Dong et al.(2014)]{DZZetal2014}
Dong, S., Zheng, Z., Zhu, Z., et al. 2014, ApJ, 789, L3
\bibitem[Gray \& Corbally(2009)]{GC2009}
Gray, R. O., \& Corbally, C. J. 2009, Stellar Spectral Classification (Princeton, NJ: Princeton Univ. Press)
\bibitem[Grigahc\`ene et al.(2010)]{GABetal2010}
Grigahc\`ene, A., Antoci, V., Balona, L., et al. 2010, ApJ, 713, L192
\bibitem[Iglesias \& Rogers(1996)]{OPAL}
Iglesias, C. A., \& Rogers, F. J. 1996, ApJ, 464, 943
\bibitem[Kallinger et al.(2008)]{KRW2008}
Kallinger, T., Reegen, P., \& Weiss, W. W. 2008, A\&A, 481, 571
\bibitem[Keen et al.(2015)]{Keen2015}
Keen, M. A., Bedding, T. R., Murphy, S. J., et al. 2015, MNRAS, 454, 1792
\bibitem[Kinemuchi et al.(2012)]{PyKE}
Kinemuchi, K., Barclay, T., Fanelli, M., et al. 2012, PASP, 124, 963
\bibitem[Koleva et al.(2009)]{KPBW2009}
Koleva, M., Prugniel, Ph., Bouchard, A., \& Wu, Y. 2009, A\&A, 501, 1269
\bibitem[Kurtz et al.(2015)]{Kurtzetal2015}
Kurtz, D. W., Shibahashi, H., Murphy, S. J., et al. 2015, MNRAS, 450, 3015
\bibitem[Lehmann et al.(2011)]{LTSetal2011}
Lehmann, H., Tkachenko, A., Semaan, T., et al. 2011, A\&A, 526, A124
\bibitem[Lenz \& Breger(2005)]{Period04}
Lenz, P., \& Breger, M. 2005, CoAst, 146, 53
\bibitem[Liu et al.(2015)]{Liuetal2015}
Liu, C., Fang, M., Wu, Y., et al. 2015, ApJ, 807, 4
\bibitem[McNamara et al.(2012)]{MNJMK2012}
McNamara, B., Jackiewicz, J., \& McKeever, J. 2012, AJ, 143, 101
\bibitem[Miglio et al. (2008)]{Miglio2008}
Miglio A., Montalb\'an J., Noels A., \& Eggenberger P. 2008, MNRAS, 386, 1487
\bibitem[Moravveji(2016)]{Moravveji2016}
Moravveji, E. 2016, MNRAS, 455, L67
\bibitem[Moravveji et al.(2015)]{Moravveji2015}
Moravveji, E., Aerts, C., P\'apics, P. I., Triana S. A., \& Vandoren, B. 2015, A\&A, 580, A27
\bibitem[Moravveji et al.(2016)]{MTAM2016}
Moravveji, E., Townsend, R. H. D., Aerts, C., \& Mathis, S. 2016, ApJ, 823, 130
\bibitem[Nielsen et al.(2013)]{NGSK2013}
Nielsen, M. B., Gizon, L., Schunker, H., \& Karoff, C. 2013, A\&A, 557, L10
\bibitem[\O stensen et al.(2010)]{Ostensen2010}
\O stensen, R. H., Silvotti, R., Charpinet, S., et al. 2010, MNRAS, 409, 1470
\bibitem[P\'apics(2012)]{Papics2012}
P\'apics, P. I. 2012, AN, 333, 1053
\bibitem[P\'apics et al.(2012)]{HD43317}
P\'apics, P. I., Briquet, M., Baglin, A. et al. 2012, A\&A, 542, A55
\bibitem[P\'apics et al.(2014)]{Papics2014}
P\'apics, P. I., Moravveji, E., Aerts, C., et al. 2014, A\&A, 570, A8
\bibitem[P\'apics et al.(2013)]{Papics2013}
P\'apics, P. I., Tkachenko, A., Aerts, C., et al. 2013, A\&A, 553, A127
\bibitem[P\'apics et al.(2015)]{Papics2015}
P\'apics, P. I., Tkachenko, A., Aerts, C., et al. 2015, ApJ, 803, L25
\bibitem[P\'apics et al.(2017)]{Papics2017}
P\'apics, P. I., Tkachenko, Van Reeth T., et al. 2017, A\&A, 598, A74
\bibitem[Paxton et al.(2015)]{MESA2015}
Paxton, B., Marchant, P., Schwab, J., et al. 2015, ApJS, 220, 15
\bibitem[Pinsonneault et al.(2012)]{PAMZetal2012}
Pinsonneault, M. H., An, D., Molenda-\.Zakowicz, J., et al. 2012, ApJS, 199, 30
\bibitem[Prugniel \& Soubiran(2001)]{ELODIE}
Prugniel, Ph., \& Soubiran, C. 2001, A\&A, 369, 1048
\bibitem[Prugniel et al.(2011)]{MILES}
Prugniel, Ph., Vauglin, I., \& Koleva, M. 2011, A\&A, 531, A165
\bibitem[Reegen(2007)]{SigSpec}
Reegen, P. 2007, A\&A, 467, 1353
\bibitem[Reinhold et al.(2013)]{RRB2013}
Reinhold, T., Reiners, A., \& Basri, G. 2013, A\&A, 560, A4
\bibitem[Ren et al.(2016)]{Ren2016}
Ren, A., Fu, J., De Cat, P., et al. 2016, ApJS, 225, 28
\bibitem[Saio et al.(2015)]{Saio2015}
Saio, H., Kurtz, D. W., Takata, M., et al. 2015, MNRAS, 447, 3264
\bibitem[Schmid et al.(2015)]{Schmid2015}
Schmid, V. S., Tkachenko, A., Aerts, C., et al. 2015, A\&A, 584, A35
\bibitem[Tassoul(1980)]{Tassoul1980}
Tassoul, M. 1980, ApJS, 43, 469
\bibitem[Tkachenko et al.(2013)]{TLSU2013}
Tkachenko, A., Lehmann, H., Smalley, B., \& Uytterhoeven, K. 2013, MNRAS, 431, 3685
\bibitem[Triana et al.(2015)]{Triana2015}
Triana, S. A., Moravveji, E., P\'apics, P. I., et al. 2015, ApJ, 810, 16
\bibitem[Ulusoy et al.(2014)]{USIU2014}
Ulusoy, C., Stateva, I., Iliev, I. Kh., \& Ula\c s, B. 2014, New Astro., 30, 28
\bibitem[Uytterhoeven et al.(2011)]{UMGetal2011}
Uytterhoeven, K., Moya, A., Grigahc\`ene, A., et al. 2011, A\&A, 534, A125
\bibitem[Van Reeth et al.(2015)]{VanReeth2015}
Van Reeth, T., Tkachenko, A., Aerts, C., et al. 2015, A\&A, 574, A17
\bibitem[Walczak et al.(2015)]{WFCKG2015}
Walczak, P., Fontes, C. J., Colgan, J., Kilcrease, D. P., \& Guzik, J. A. 2015, A\&A, 580, L9
\bibitem[Wu et al.(2014)]{Wu2014}
Wu, Y., Du, B., Luo A., et al. 2014, in Proc. IAU Symp. 306, Statistical Challenges in 21st Century Cosmology, ed. A. F. Heavens, J.-L. Starck, \& A. Krone-Martins (Cambridge: Cambridge Univ. Press), 340
\bibitem[Wu et al.(2011a)]{Lpipeline}
Wu, Y., Luo, A., Li, H., et al. 2011a, RAA, 11, 924
\bibitem[Wu et al.(2011b)]{WSPGK2011}
Wu, Y., Singh, H. P., Prugniel, P., et al. 2011b, A\&A, 525, A71
\bibitem[Zhao et al.(2012)]{Lsurvey}
Zhao, G., Zhao, Y., Chu, Y., et al. 2012, RAA, 12, 723
\bibitem[Zwintz et al.(2017)]{Zwintz2017}
Zwintz, K., Moravveji, E., P\'apics, P. I., et al. 2017, A\&A, 601, A101
\end{thebibliography}
\end{document}